\documentclass[fleqn,usenatbib]{mnras}

\usepackage{newtxtext,newtxmath}

\usepackage[T1]{fontenc}

\DeclareRobustCommand{\VAN}[3]{#2}
\let\VANthebibliography\thebibliography
\def\thebibliography{\DeclareRobustCommand{\VAN}[3]{##3}\VANthebibliography}

\usepackage{graphicx}	
\usepackage{amsmath}	

\title[Turbulence modes in M33]{Turbulence modes in molecular clouds in M33: a challenging test of the Brunt-Federrath framework}

\author[R. Rani et al.]{
Raffaele Rani$^{1}$\thanks{E-mail: yuwei@kurima.art (RR)},
Ayu Konishi$^{2,3}$,
Kazuyuki Muraoka$^{2}$,
Toby J. T. Moore$^{4}$,
David J. Eden$^{5,6}$,
\newauthor
\  Karan Fernandes $^{7,8}$,
Yueh-Ning Lee  $^{9}$,
Shih-Ping Lai $^{1}$
\\
\\
$^{1}$Institute of Astronomy, National Tsing Hua University, General Building II, NTHU, No. 101, Section 2, Kuang-Fu Road, Hsinchu 30013, Taiwan, R.O.C \\
$^{2}$Department of Physics, Graduate School of Science, Osaka Metropolitan
University, 3-3-138 Sugimoto, Sumiyoshi-ku, Osaka, 558-8585, Japan\\
$^{3}$Astronomical Institute, Tohoku University, 6-3, Aramaki, Aoba-ku, Sendai, Miyagi 980-8578, Japan \\
$^{4}$Astrophysics Research Institute, Liverpool John Moores University, IC2, Liverpool Science Park, 146 Brownlow Hill, Liverpool L3 5RF, UK\\
$^{5}$Global Banking School, Universal Square, Devonshire Street North, Manchester, M12 6JH, UK\\
$^{6}$Armagh Observatory and Planetarium, College Hill, Armagh, BT61 9DB, UK\\
$^{7}$Subatomic Physics Research Group, Science and Technology Advanced Institute,
Van Lang University, 69/68 Dang Thuy Tram Street, Binh Loi Trung Ward,\\
Ho Chi Minh City 70000, Vietnam\\
$^{8}$Faculty of Applied Technology, Van Lang School of Technology,
Van Lang University, 69/68 Dang Thuy Tram Street, Binh Loi Trung Ward,\\
Ho Chi Minh City 70000, Vietnam\\
$^{9}$Center of Astronomy and Gravitation, National Taiwan Normal University, 88, Sec. 4, Ting-Chou Rd., Wenshan District, Taipei 116, Taiwan R.O.C.
}

\date{Accepted XXX. Received YYY; in original form ZZZ}

\pubyear{\the\year{}}

\begin{document}
\label{firstpage}
\pagerange{\pageref{firstpage}-\pageref{lastpage}}
\maketitle

\begin{abstract}
Understanding the turbulent structure of molecular clouds is essential for linking galactic environment to the regulation of star formation. Although Galactic molecular clouds have been studied extensively, extending turbulence-mode analysis beyond the Milky Way remains challenging because of the limited spatial resolution and dynamic range of extragalactic observations. We assess the applicability of the Brunt \& Federrath (2014; BF14) framework to molecular clouds in M33. We apply BF14 to 128 $^{13}$CO(2-1) clouds, selecting a primary quality-controlled sample of 54 clouds and a more restrictive
25-cloud subsample defined using a stricter extraction-padding
criterion (PAD=2) for quantitative analysis.
 Tests of cloud and noise power spectra, Fourier-space cutoffs, extraction padding, and Monte Carlo uncertainties demonstrate that the retained spatial frequencies are dominated by astrophysical signal and that the inferred turbulence-mode fractions are robust against reasonable analysis choices. 
Both samples yield consistent distributions of the solenoidal fraction,
with mean values of $R \simeq 0.22$ for both samples and median values of
$R \simeq 0.18$ and $0.17$, respectively, indicating that compressive
motions dominate the turbulent energy budget. No strong systematic variations of $R$ are found with galactocentric distance or tracers of recent star formation. These results demonstrate that, despite the challenging spatial resolution of current extragalactic molecular-line observations, the BF14 framework can be applied reliably to nearby galaxies using spectroscopic observations alone. To our knowledge, this constitutes the first application of the full BF14 formalism to extragalactic molecular-cloud observations.
\end{abstract}




\section{Introduction}

\begin{figure}
    \centering
    \includegraphics[width=\columnwidth, 
    height=0.8\textheight]{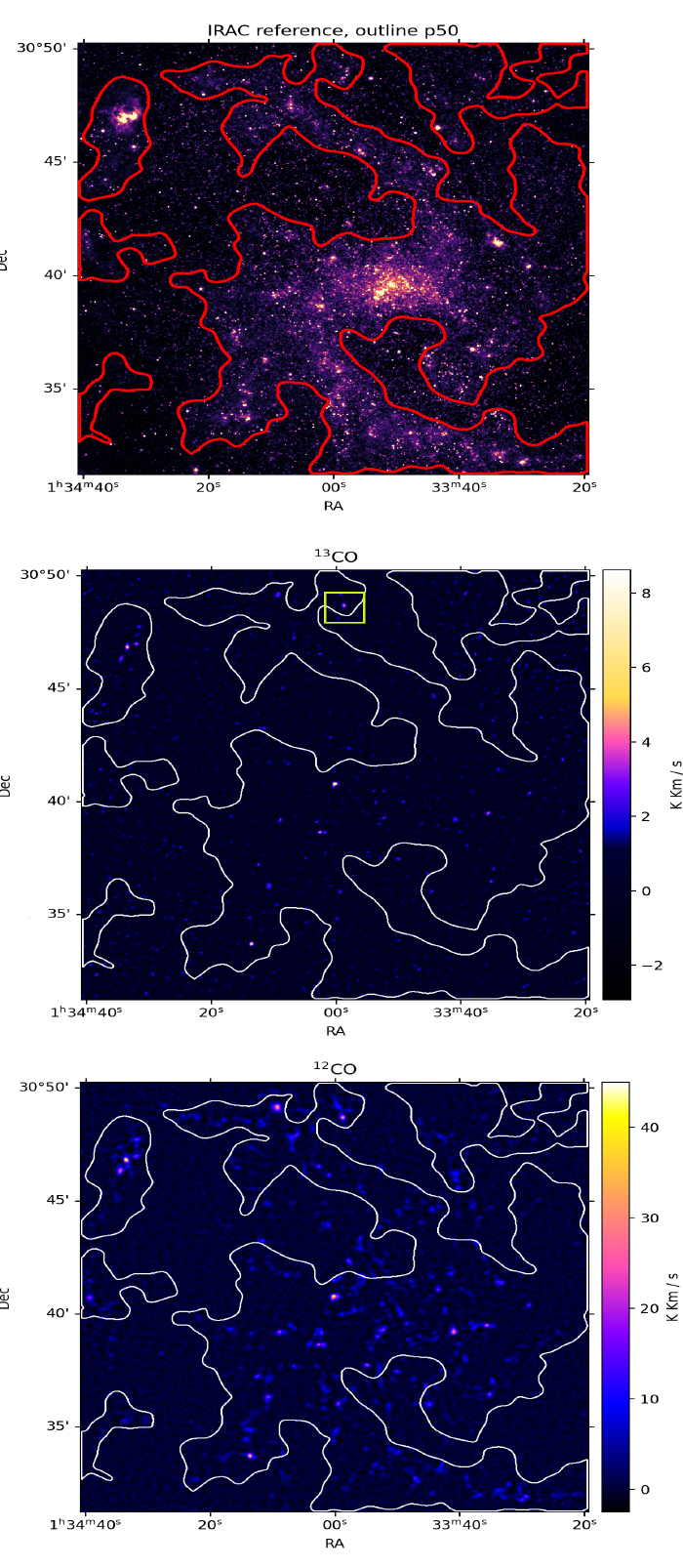}
    \caption{Comparison between the large-scale stellar structure of M33 and the molecular gas traced by CO emission. \textit{Left:} IRAC $3.6 \mu$m map used as a reference for the stellar distribution. The image was smoothed with a Gaussian kernel ($\sigma = 12$ pixels) and processed with an unsharp-masking procedure to enhance the large-scale structure of the disc. The red contour shows the 50th percentile intensity contour of the arm-enhanced map, outlining the global stellar disc morphology.  \textit{Middle:} $^{13}$CO($2\!-\!1$) integrated intensity map.  \textit{Right:} $^{12}$CO($2\!-\!1$) integrated intensity map. In the $^{13}$CO and $^{12}$CO panels, the white contour corresponds to the same stellar reference boundary derived from the IRAC map (left panel). The yellow rectangle in the middle panel shows the representative molecular cloud (ID 1177), which is used as the reference cloud in the figures below.}
    \label{emission}
\end{figure}


Turbulent motions are observed across nearly all astrophysical
environments and play a central role in shaping the dynamical evolution
of gaseous systems, from stellar interiors \citep{Miesch2009} to the
intracluster medium of galaxy clusters \citep{Zhuravleva2014}.
Within the interstellar medium (ISM), turbulence governs the structure
and evolution of molecular gas, mediating the interplay between
gravity, stellar feedback, magnetic fields and large-scale gas flows
\citep{Burkhart2020,Rosen2020,Girichidis2020}.
As a consequence, turbulence represents one of the primary regulators
of star formation in galaxies.

An important characteristic of interstellar turbulence is the balance
between compressive and solenoidal motions. Numerical studies have
shown that different turbulence modes lead to markedly different
evolutionary outcomes for molecular clouds.
Turbulence dominated by compressive (curl-free) motions efficiently
concentrates gas into dense regions, thereby enhancing gravitational
collapse, whereas solenoidal (divergence-free) motions redistribute
kinetic energy through rotational and shear flows that oppose
large-scale compression
\citep{Federrath2008,Federrath2010,FederrathKlessen2012}.
The relative balance between these modes therefore influences the
density structure of molecular clouds and ultimately their star
formation efficiency
\citep{Krumholz2005,Padoan2014, FederrathKlessen2013,Kainulainen2013,Burkhart2015}.

In turbulent molecular clouds, the statistical properties of the
density field are closely connected to the nature of the turbulent
motions. Compressive turbulence produces broader density probability
distribution functions and enhances the formation of high-density
structures, whereas solenoidal turbulence tends to maintain a narrower
density distribution
\citep{Federrath2010,FederrathKlessen2012}.
As a result, variations in the balance between these turbulence modes
can lead to substantial differences in the amount of gas available for
gravitational collapse and star formation. In the Brunt \& Federrath (2014) framework, this balance is quantified
by the solenoidal momentum fraction,
$R \equiv \langle p_\perp^2\rangle/\langle p^2\rangle$, where
$p_\perp$ is the divergence-free component of the turbulent momentum
density. Larger values of $R$ therefore correspond to a greater
contribution from solenoidal motions, whereas smaller values indicate
a more compressive velocity field. This quantity is conceptually
related to the turbulence-driving parameter $b$, commonly inferred
from the density-variance-Mach-number relation
\citep[e.g.][]{FederrathKlessen2012}: both characterize the relative
importance of compressive and solenoidal turbulence, although they are
not equivalent quantities. The parameter $b$ describes the
compressibility or forcing character of the turbulence, whereas $R$
directly measures the relative solenoidal contribution to the
turbulent momentum field.

Observational constraints on turbulence modes have only recently
become accessible. Measurements within Galactic molecular clouds
indicate that the fraction of solenoidal and compressive motions varies
both between clouds and within individual regions.
For instance, predominantly solenoidal turbulence has been linked to
relatively inefficient star formation in nearby clouds such as Orion B
\citep{Orkisz2017}, while \citet{Federrath2016} inferred strongly
solenoidal driving in the Central Molecular Zone cloud G0.253+0.016
(the Brick), with $b=0.22\pm0.12$. Analyses of larger Galactic samples
also reveal systematic variations of turbulence properties across the
Milky Way disc \citep{Rani2022}.
These findings suggest that variations in turbulence modes may
contribute to the diversity of star formation efficiencies observed
among molecular clouds \citep{Eden2013,Eden2015}.
Whether similar behaviour persists beyond the Milky Way remains
largely unknown. Extending turbulence studies to external galaxies is
essential for disentangling the relative importance of local
cloud-scale physics and the global galactic environment. However,
observational investigations of turbulence modes in extragalactic
systems remain scarce and are typically restricted to individual
regions where detailed multi-wavelength datasets permit the
reconstruction of gas density probability distribution functions
(PDFs; \citealt{Sharda2018,Schneider2015,Menon2021,Sharda2021}).
Such approaches require extensive observational coverage and are
challenging to apply systematically across entire molecular cloud
populations.
A complementary approach is provided by the statistical formalism
developed by \citet{Brunt2010,Brunt2014}, which enables the estimation
of the relative power contained in solenoidal turbulent motions using
line-of-sight observables alone. This technique circumvents the need
for density-PDF measurements and has recently been successfully
implemented for Galactic molecular-cloud surveys \citep{Rani2022}.
That study also found a significant negative correlation between the
solenoidal fraction and the estimated star formation efficiencies of
Galactic-plane molecular clouds, indicating that a larger contribution
from compressive motions is associated with more efficient star formation. Because it relies primarily on spectroscopic
observations, the method is particularly well suited to nearby galaxies
observed with modern interferometric facilities.
Whether the Brunt-Federrath (BF14) framework can be applied reliably to
molecular clouds beyond the Milky Way, however, remains largely
unexplored. An extragalactic application for the determination of a turbulence-driving factor based on \cite{Brunt2010} was presented by \citet{Gerrard2023}, who extended the formalism to produce spatially resolved maps of the turbulence-driving mixture in
H\,\textsc{i} observations of the Small Magellanic Cloud.
Observational applications of the full BF14 to extragalactic molecular clouds remain absent, where
the limited spatial resolution and dynamic range of extragalactic CO observations present significant challenges for recovering the turbulent power spectrum required by the method. However, to our knowledge, the full BF14 method has yet to be tested at the resolution of extragalactic  applications.  
Establishing the applicability of the Brunt-Federrath framework to
molecular clouds in nearby galaxies would enable direct comparisons of
turbulence modes across different galactic environments using
spectroscopic observations alone, making the method particularly
attractive for future large extragalactic molecular cloud surveys.

Despite its observational simplicity, the Brunt-Federrath framework
places stringent requirements on the statistical properties of the
observed emission.
The method relies on an accurate reconstruction of the spatial power
spectrum over a sufficiently broad range of Fourier modes, making it
potentially sensitive to limited spatial resolution, beam convolution,
finite cloud size, extraction boundaries, and observational noise.
These effects become increasingly important for extragalactic
molecular cloud observations, where individual clouds are only
marginally resolved and the available dynamic range in spatial scales
is substantially smaller than in Galactic surveys.
Consequently, before the inferred turbulence modes can be interpreted
physically, it is essential to establish that the underlying
assumptions of the Brunt-Federrath formalism remain valid under these
observational conditions.
The Triangulum galaxy (M33) provides a valuable test case for extending
the Brunt-Federrath framework beyond the Milky Way. As a nearby Local
Group galaxy located at a distance of approximately 840\,kpc and viewed
at relatively low inclination, M33 offers a resolved view of molecular
clouds across much of its disc. Recent Atacama Compact Array (ACA) and
Atacama Large Millimeter/submillimeter Array (ALMA) observations have
enabled the identification and characterisation of a large, homogeneous
sample of giant molecular clouds in $^{13}$CO(2-1)
\citep{Muraoka2023}. At the corresponding physical resolution of
$\sim30$\,pc, individual clouds are only marginally resolved, placing
the Brunt-Federrath analysis close to the practical resolution limit
of current extragalactic molecular-line observations. In addition to the spatial-resolution limitation, the parent
ACA+IRAM $^{12}$CO cloud catalogue has an estimated completeness
limit of $M_{\rm CO}\simeq5\times10^{4}\,M_\odot$
\citep{Muraoka2023}. The requirement of significant $^{13}$CO
emission imposes a further surface-brightness and column-density
selection, preferentially retaining the denser and higher-$\Sigma_{\rm H_2}$
parts of the molecular-cloud population. The resulting BF14 sample
should therefore not be regarded as complete toward low-mass or
low-surface-density clouds.
The combination
of a uniform cloud catalogue and broad environmental coverage therefore
makes M33 a particularly useful benchmark for assessing the robustness
of the method under extragalactic observing conditions.

M33 is a flocculent late-type spiral galaxy, classified as SA(s)cd, with
a lower stellar mass, weaker spiral structure, reduced large-scale shear,
and lower metallicity than the Milky Way. It has a stellar mass of
approximately $5\times10^9\,M_\odot$ and a characteristic metallicity of
about $0.5\,Z_\odot$ \citep{Corbelli2003,RosolowskySimon2008}. These
differences in global galactic structure and interstellar-medium
conditions may influence the balance between compressive and solenoidal
turbulent motions through their effects on cloud assembly, large-scale
gas flows, gravitational instability, and the thermal and dynamical
state of the gas.  Published estimates also suggest that M33 may have a relatively high
molecular-gas star formation efficiency compared with more massive
spiral galaxies, providing an additional reason to investigate whether
its molecular clouds exhibit a systematically different balance of
compressive and solenoidal turbulent motions
\citep{Gardan2007,Gratier2010,Kang2012}.

In this work, we present a challenging test of the Brunt-Federrath
framework by applying it to a sample of $^{13}$CO(2-1) molecular
clouds in M33.
Using two independently selected cloud samples, we investigate the
effects of spatial resolution, Fourier-space sampling, extraction
padding, beam cutoff, and observational noise on the recovered
solenoidal fraction.
Having established the robustness of the method under these
conditions, we characterise the distribution of turbulence modes
within the molecular cloud population of M33 and examine their
relationship with tracers of recent star formation.

The article is organised as follows. Section 2 describes the
observational data. Section 3 describes the cloud identification.
Section 4 outlines the methodology used to estimate the solenoidal
fraction. Section 5 presents the results. Section 6 discusses the
implications for turbulence modes and star formation, and Section 7
summarises the main conclusions.

\section{Data}\label{data}
Our primary data set consists of $^{13}$CO(2-1) observations,
together with complementary $^{12}$CO(2-1) and C$^{18}$O(2-1)
data acquired with the ACA 7-m antennas of ALMA. The ACA
$^{12}$CO(2-1) data cube was combined with IRAM 30-m
observations \citep{Druard2014} to recover extended molecular-line
emission; hereafter, we refer to this product as the
ACA+IRAM $^{12}$CO(2-1) data cube.

The $^{12}$CO, $^{13}$CO, and C$^{18}$O data have synthesized beam
sizes of $7.31\times6.50$ arcsec, $7.72\times6.86$ arcsec, and
$7.82\times6.96$ arcsec, corresponding to physical resolutions of
approximately $30\times26$ pc, $31\times27$ pc, and $31\times28$ pc,
respectively. The corresponding velocity resolutions are 0.7, 1.4,
and 1.6 km s$^{-1}$. The corresponding rms noise levels are 39, 30, and 22 mK for
$^{12}$CO, $^{13}$CO, and C$^{18}$O, respectively. Full details of the observing strategy
and data-reduction pipeline are presented by \citet{Muraoka2023}.

Figure~1 shows the integrated-intensity maps of $^{13}$CO from the
standalone ACA data and $^{12}$CO from the combined ACA+IRAM data. Following \citet{Muraoka2023}, the integrated-intensity maps
are constructed by integrating each spatial pixel over a velocity
interval of $V_{\rm rep}\pm30~{\rm km\,s^{-1}}$, where $V_{\rm rep}$
is the representative local H\,{\sc i} velocity derived from the
H\,{\sc i} velocity field. No additional intensity-threshold mask is
applied to the moment maps; restricting the integration to this
velocity interval limits the contribution from line-free noise.
Both maps trace the molecular-gas distribution across M33 at
approximately 30-pc resolution, while the addition of the IRAM 30-m
data enables the $^{12}$CO map to recover more extended molecular
emission.
The integrated $^{13}$CO emission across the M33 disc exhibits many
features in common with the $^{12}$CO emission, tracing moderately
dense gas with characteristic densities of $\sim10^3$~cm$^{-3}$
within the $^{12}$CO clouds \citep{Muraoka2023}. We note, however,
that the two data sets are not strictly matched in their spatial
filtering: the $^{12}$CO cube includes the IRAM 30-m single-dish
component, whereas the $^{13}$CO observations contain ACA data only.
The combined $^{12}$CO data therefore recover more extended emission
than the $^{13}$CO cube. This difference is not expected to strongly
affect the identification of compact $^{13}$CO-bright structures,
since diffuse molecular emission is substantially fainter in
$^{13}$CO than in $^{12}$CO.

We construct an H$_2$ column-density map from the integrated
$^{12}$CO emission using the CO-to-H$_2$ conversion factor,
$X_{\rm CO}$, and the line ratio $R_{2-1/1-0}$, defined as the
$^{12}$CO(2-1)/$^{12}$CO(1-0) integrated-intensity ratio. We adopt
the M33-specific values
$X_{\rm CO}=4.0\times10^{20}\,
\mathrm{cm^{-2}\,(K\,km\,s^{-1})^{-1}}$
\citep{Gratier2017} and $R_{2-1/1-0}=0.60$ throughout the M33 disc,
following the prescription adopted by \citet{Muraoka2023}.

To highlight the spiral arm structure in M33, we use
\textit{Spitzer} IRAC 3.6-$\mu$m near-infrared data.
Emission at this wavelength primarily traces the old stellar
population, which dominates the stellar mass of disc galaxies
\citep{McQuinn2007}.
Following \citet{Konishi2024}, the spiral features were enhanced
by subtracting an axisymmetric model of the surface brightness
\citep{Regan1994}. The resulting map is also used in
Figure \ref{emission} to define the stellar-reference contours
overlaid on the CO integrated-intensity maps.

To account for star formation activity within the sample of
molecular clouds extracted from the ACA $^{13}$CO map, we use both
far-infrared imaging of M33 and a catalogue of 70-$\mu$m point
sources.
Imaging at 70\,$\mu$m was obtained with the Photodetector Array
Camera and Spectrometer (PACS) aboard \textit{Herschel} as part
of the Herschel M33 Extended Survey (HerM33ES;
\citealt{Kramer2010}) and accessed through the
\textit{Herschel} Science Archive.
The PACS point-source catalogues used in this work are taken from
the Herschel/PACS Point Source Catalogue \citep{Marton2024}. The PACS $70\,\mu$m maps have a point-spread function (PSF)
with FWHM $\simeq 5.6$ arcsec, corresponding to $\simeq 22$ pc
at the distance of M33, and are sampled at 2 arcsec per pixel,
such that one PSF area covers approximately nine pixels
\citep{Poglitsch2010}.
The 70-$\mu$m emission is commonly used as a tracer of recent star
formation, as it probes dust heated by young massive stars and their
surrounding environments.
Given the modest angular resolution of the PACS 70-$\mu$m data,
extended emission from H\,{\sc ii} regions and their surrounding
photodissociation regions is expected to dominate over compact
embedded sources. These data therefore primarily trace recently
formed massive stars and their immediate environments on
timescales of $\sim 10^5$-$10^6$\,yr.
To compare the locations of the $^{13}$CO sources with sites of
recent star formation in M33, we additionally
use a 5-GHz radio continuum map obtained with the
\textit{Karl G. Jansky Very Large Array} (VLA)
\citep{White2019}.  The 5-GHz radio-continuum map has an angular resolution of
$5.9$~arcsec FWHM, corresponding to a physical scale of approximately
24~pc at the distance of M33 \citep{White2019}. This is closely matched
to the $\sim30$~pc resolution of the molecular-line observations. Radio continuum emission at 5\,GHz traces a combination of thermal
free-free emission from H{\sc ii} regions powered by massive,
short-lived stars and non-thermal synchrotron emission from
cosmic-ray electrons accelerated by supernovae.
At this frequency, the relative contribution of the thermal
component is enhanced compared to lower radio frequencies,
making 5-GHz emission a particularly effective,
extinction-free tracer of star formation on timescales of
$\sim$\,Myr \citep{Murphy2011}.

\section{Cloud identification methods}\label{methods}

\subsection{Segmentation of the $^{13}$CO map - PYCPROPS}

The molecular gas distribution across the disc of M33 exhibits a
complex and hierarchical structure. In the following analysis,
$^{13}$CO emission is used because of its substantially lower optical
depth compared to $^{12}$CO. The current Brunt--Federrath formalism
requires an approximately optically thin tracer, such that the observed
line intensity preserves the density weighting assumed in the
construction of the moment fields. Although the brighter $^{12}$CO
emission would provide a substantially larger sample of molecular
clouds, its typically high optical depth violates this requirement and
therefore prevents a direct application of the BF14 method. We
therefore use $^{13}$CO, prioritising compliance with the assumptions
of the formalism over the larger statistics obtainable with $^{12}$CO.
As shown in Section~4.1, the $^{13}$CO optical depths derived for the
M33 data remain sufficiently low for the application of the method. We employ {\sc pycprops} \citep{Rosolowsky2021}, a Python
implementation of the {\sc cprops} algorithm
\citep{Rosolowsky2006}, to segment the $^{13}$CO emission and
construct a molecular-cloud catalogue.

The algorithm first identifies contiguous regions of statistically
significant emission in position-position-velocity (PPV) space by
applying signal-to-noise thresholds to the data cube. Voxels exceeding
the specified high- and low-significance thresholds are grouped into
connected structures using a three-dimensional connectivity criterion,
ensuring that detected features are spatially and kinematically
coherent. This masking step suppresses noise-dominated voxels and
defines the overall extent of molecular emission considered for cloud
identification.

Within the masked emission, {\sc pycprops} identifies individual
clouds by locating local maxima in intensity and assigning surrounding
emission to these peaks through a watershed-style partitioning of the
PPV volume. Each cloud is thus defined as a contiguous set of voxels
associated with a single emission maximum, without assuming any
predefined geometry or functional form for the emission profile. This
approach allows the algorithm to capture the irregular and
non-spherical morphologies commonly observed in molecular gas.

For each identified cloud, {\sc pycprops} computes a suite of
physical and kinematic properties, including the flux-weighted
centroid position, effective radius, velocity dispersion, and
integrated intensity.
We follow the implementation described by \citet{Muraoka2023}, with
adjustments tailored to the $^{13}$CO data. Before cloud
identification, the $^{13}$CO cube is spectrally resampled onto the
same velocity grid as the $^{12}$CO cube, with a channel spacing of
0.7\,km~s$^{-1}$. Placing the two data sets on a common velocity grid
facilitates the subsequent association of the $^{13}$CO clouds with
the $^{12}$CO cloud catalogue and ensures that compatible
channel-based masking and segmentation criteria are applied to both
data sets. The resampling is performed before cloud extraction
because altering the velocity sampling afterwards could change the
connectivity of the significant PPV voxels and hence the shapes and
boundaries of the identified clouds. We note that this operation does
not improve the intrinsic spectral resolution of the $^{13}$CO data,
which remains 1.4\,km~s$^{-1}$.

High- and low-significance masking thresholds are set to $4\sigma$
and $3\sigma$, respectively \citep{Rosolowsky2021}. In this scheme,
voxels exceeding the $4\sigma$ threshold define robust emission
peaks, while spatially and spectrally connected voxels above
$3\sigma$ are included to recover the surrounding lower-intensity
emission. Significant emission is required to extend across at least
three consecutive channels in the regridded cube and to cover a
projected area of at least one synthesized beam. For the $^{13}$CO
data, the beam is $7.72\times6.86$\,arcsec, corresponding to
approximately $31\times27$\,pc at the distance of M33. These criteria
define the minimum extent of significant emission used to construct
the initial PPV islands and suppress spurious noise detections; they
do not impose a minimum size on the individual molecular clouds
identified by the subsequent watershed segmentation.

Local maxima are then required to be separated by at least one beam in
the spatial dimension and by $7$\,km\,s$^{-1}$ in velocity space,
following the procedure adopted by \citet{Muraoka2023}. This minimum
spectral separation corresponds approximately to the characteristic
velocity width of a $\sim30$-pc giant molecular cloud inferred from
the Galactic size-linewidth relation
($\sigma_v = 0.72\,r^{0.5}$; \citealt{Solomon1987}),
and therefore
prevents the algorithm from artificially splitting emission
structures whose velocity extent is consistent with that expected for
a single cloud. These parameter choices ensure that the identified
structures correspond to molecular clouds resolved at the $\sim30$-pc
spatial scale of the observations. We note that the segmentation is performed independently on both the
$^{13}$CO and $^{12}$CO cubes. For the latter, we repeat the
{\sc pycprops} extraction using the parameters adopted by
\citet{Muraoka2023}, rather than using their published cloud
catalogue.
To ensure that the molecular-cloud sample is suitable for the
subsequent Brunt--Federrath analysis, we additionally require each
$^{13}$CO assignment to span at least $4\times4\times2$ voxels in PPV
space. Given the spatial sampling of 1.5 arcsec pixel$^{-1}$ and the
0.7~km~s$^{-1}$ channel spacing of the regridded cube, this corresponds
to a minimum extent of $6\times6$ arcsec spatially and
1.4~km~s$^{-1}$ in velocity. This criterion is therefore tied to the
physical sampling of the observations rather than to an arbitrary
voxel count; finer numerical regridding would not increase the
intrinsic spatial or spectral resolution of the data.
Applying this initial selection produces a sample of 175 $^{13}$CO sources. More
restrictive quality criteria based on the number of usable Fourier
modes are applied later when defining the final Brunt--Federrath
samples.

We then compare the resulting $^{13}$CO segmentation with an
independently derived segmentation of the ACA+IRAM $^{12}$CO cube.
The $^{12}$CO extraction was repeated using the \textsc{pycprops}
parameters adopted by \citet{Muraoka2023}, rather than using their
published cloud catalogue. Only $^{13}$CO clouds that overlap
partially or completely with a $^{12}$CO structure in full PPV space,
sharing at least one voxel in position-position-velocity coordinates,
are retained. This association reduces the sample from 175 to 162 individual molecular clouds. 

To verify that the observed $^{13}$CO and $^{12}$CO association is not produced by chance, we performed a Monte Carlo experiment in which the pre-overlap $^{13}$CO assignment cube 
was randomly shifted in PPV space
while preserving its internal structure. This procedure was repeated
500 times to construct a null distribution of expected overlaps.
We find that the observed number of overlapping clouds
($N_{\rm obs} = 162$) is significantly higher than the random expectation ($\langle N_{\rm rand} \rangle = 4.9$ with a standard deviation of $5.6$).
Similarly, the number of overlapping voxels exceeds the random
expectation by more than two orders of magnitude
($2.2\times10^{4}$ compared to $\sim2\times10^{2}$).
None of the random realizations reproduces the observed level of
overlap, implying a probability $p \ll 10^{-3}$ that the association
arises from chance alignment. We therefore conclude that the spatial
coincidence between $^{13}$CO and $^{12}$CO structures is highly
significant and reflects a genuine physical association rather than a
selection bias. The expected random coincidence rate is therefore only
$\sim4$ per cent of the observed overlap.
This means that, on average, only about five of the 162 matched clouds would be expected to overlap by chance.

Finally, to ensure compatibility with the column-density map required
for the Brunt-Federrath analysis (Section ~\ref{solenoidal_fraction}), we
exclude from the {\sc PyCprops} assignment cube and catalogue all
clouds whose projected footprints overlap in the plane of the sky.
This guarantees that each molecular cloud corresponds to an isolated
column-density structure, avoiding line-of-sight blending and
boundary effects that could bias the Fourier-based analysis.
The resulting catalogue contains 128 molecular clouds and forms the
parent sample from which the quality-controlled samples used in the
subsequent analysis are selected.

\subsection{Infrared emission - Photutils}

To identify extended star-forming structures in the PACS 70-$\mu$m maps, a segmentation analysis is performed using the {\sc photutils} package \citep{Bradley2022}. The analysis operates directly on the two-dimensional surface-brightness map and aims to isolate spatially coherent regions of significant far-infrared emission.

The segmentation procedure begins by applying a surface-brightness threshold to the image, selecting pixels with flux densities exceeding a user-defined detection level above the local background. These pixels are then grouped into individual segments using a connected-component labeling algorithm such that each segment represents a contiguous region of emission connected through adjacent pixels. This approach is analogous to classical image-segmentation techniques widely used in astronomical source detection, but without imposing assumptions about source morphology or symmetry.
To ensure that the identified segments correspond to spatially
resolved emission structures, a minimum segment area of nine pixels is
imposed. Given the 2-arcsec pixel sampling, this corresponds to an area
of 36~arcsec$^2$, equivalent to a characteristic linear scale of
approximately 6-7~arcsec, or $\sim25$-30~pc at the distance of M33.
This is comparable to the PACS 70-$\mu$m beam and suppresses isolated
noise fluctuations and structures smaller than the instrumental
resolution.

The segmentation is used only to construct an assignment map of the
extended 70-$\mu$m emission for comparison with the molecular-cloud
positions; no independent photometric catalogue of infrared sources is
derived.

As a result, compact emission associated with individual embedded
YSOs is not reliably isolated by this method. The retained segments
therefore preferentially trace extended emission associated with
H\,{\sc ii} regions and surrounding photodissociation regions
(PDRs), corresponding to somewhat more evolved stages of star
formation.
This segmentation-based approach provides a complementary view to point-source catalogues, as it captures the morphology and spatial extent of extended star-forming complexes. As such, it is particularly well suited for comparison with the spatial distribution of molecular
clouds, whose sizes are typically comparable to or larger than the PACS beam, and for assessing the relationship between cloud-scale turbulence properties and star formation traced on $\gtrsim$Myr timescales.

\begin{figure*}
	\includegraphics[width=0.5\textwidth]{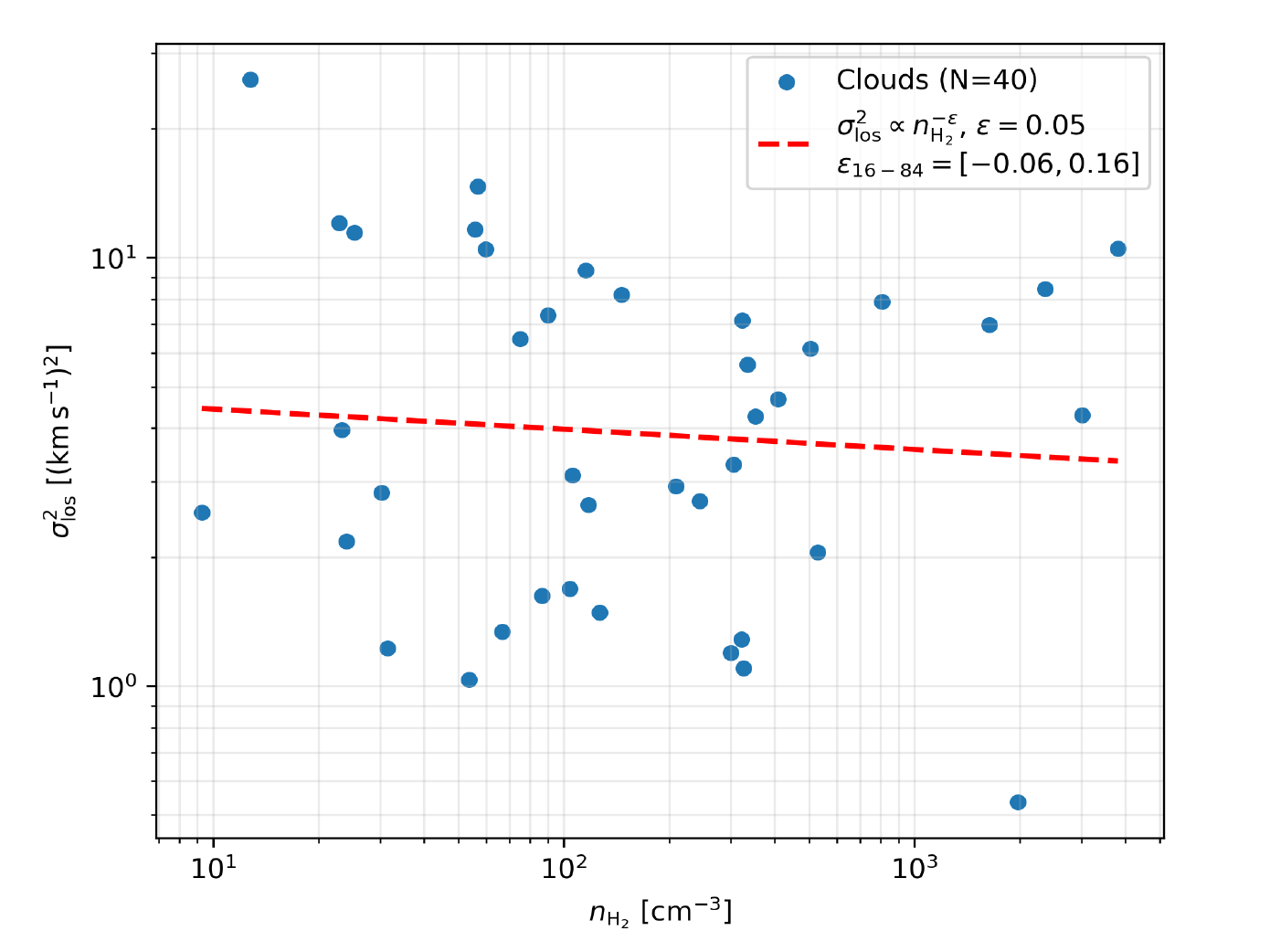}
\caption{
Scatter plot of the line-of-sight velocity variance,
$\sigma_{\rm los}^2$, as a function of the molecular-hydrogen
volume density, $n_{\rm H_2}$, for the resolved GMC subsample
($N=40$). The cloud sizes used to calculate $n_{\rm H_2}$ are
derived from beam-deconvolved spatial second moments measured
with \textsc{pycprops}, while $\sigma_{\rm los}$ is corrected for
the finite spectral resolution. The red dashed line shows the best-fitting relation
$\sigma_{\rm los}^2\propto n_{\rm H_2}^{-\epsilon}$,
with $\epsilon=0.05$. The 16th-84th percentile bootstrap
interval is $-0.06<\epsilon<0.16$, consistent with no
significant dependence of velocity variance on cloud density.}

	\label{sig_den}
\end{figure*}

\begin{figure*}
	\includegraphics[width=1.0\textwidth]{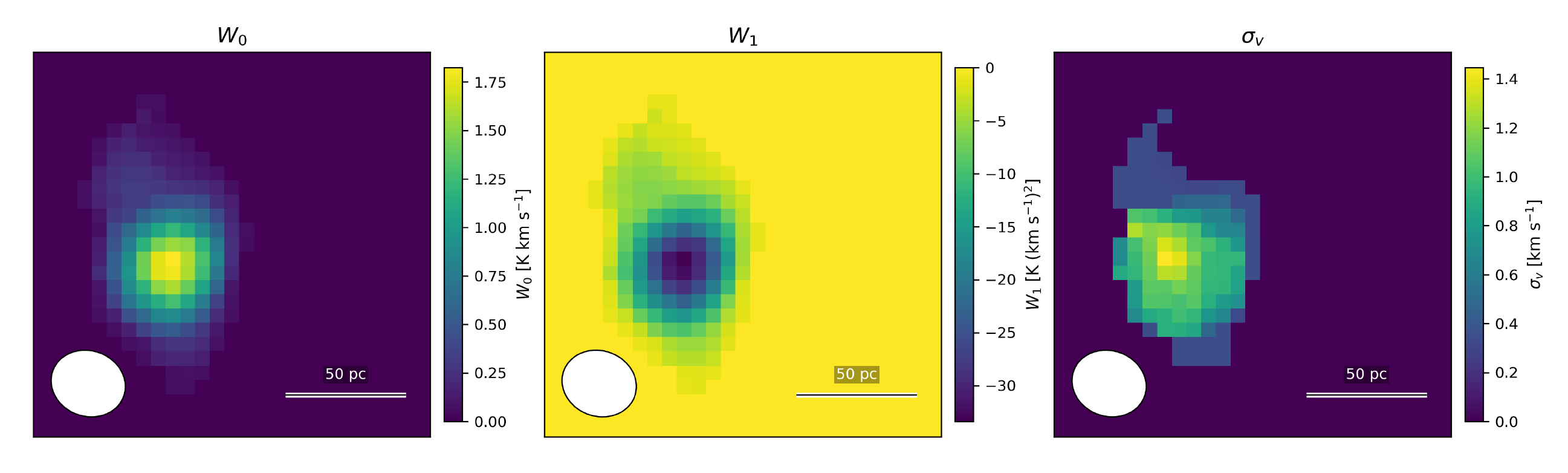}
\caption{
Moment maps of cloud 1177, selected as a representative molecular
cloud from the analysed sample and used for the methodological tests
in the following figures. The maps show the extraction region with
PAD=3 (see Section~4.5).
Left: integrated-intensity ($W_0$) map.
Centre: intensity-weighted velocity-centroid map.
Right: intensity-weighted velocity-dispersion (second-moment) map.
Colour bars give the corresponding physical units. The ellipse in the
lower-left corner of each panel shows the $7.72\times6.86$ arcsec
$^{13}$CO synthesized beam, while the scale bar corresponds to
50 pc at the distance of M33.}

	\label{cloud_1177}
\end{figure*}

\begin{figure*}
	\includegraphics[width=1.0\textwidth]{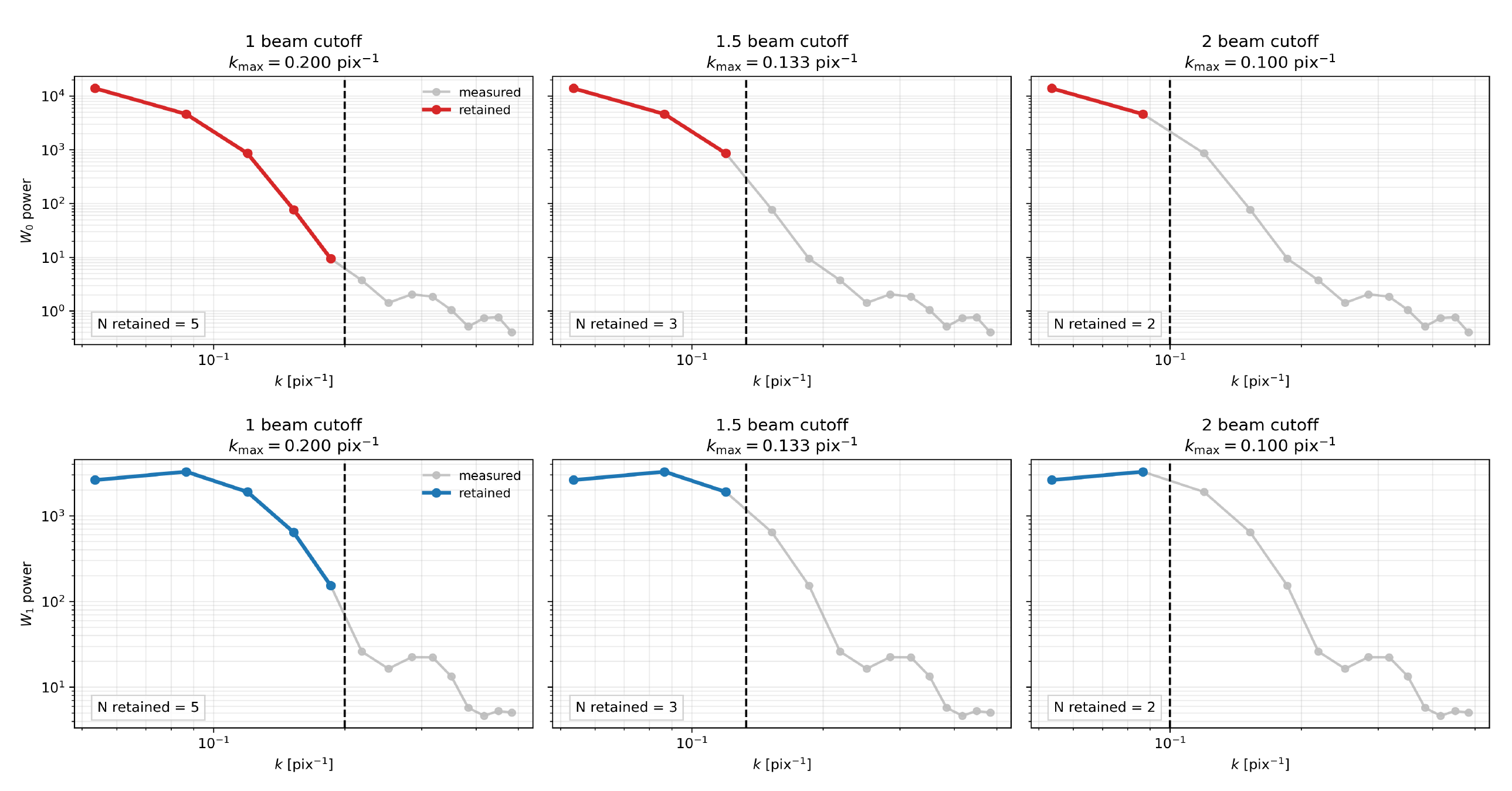}
	\caption{
Comparison of the spatial-frequency cutoffs adopted for the Brunt-Federrath analysis using cloud~1177, selected as a representative molecular cloud from the analysed sample.
The top row shows the integrated intensity ($W_0$) power spectrum, while the bottom row shows the velocity centroid ($W_1$) power spectrum.
The three columns correspond to cutoff scales of one beam (left), 1.5 beams (centre), and two beams (right).
The grey curves show the complete measured power spectra, while the coloured segments indicate the Fourier modes retained for the calculation of the solenoidal fraction.
The vertical dashed line marks the adopted cutoff spatial frequency in each case, and the number of retained Fourier bins is indicated in the lower-left corner of each panel.
}
\label{cutoff_fig}
\end{figure*}

\begin{figure*}
    \centering
    \includegraphics[width=\textwidth]
    {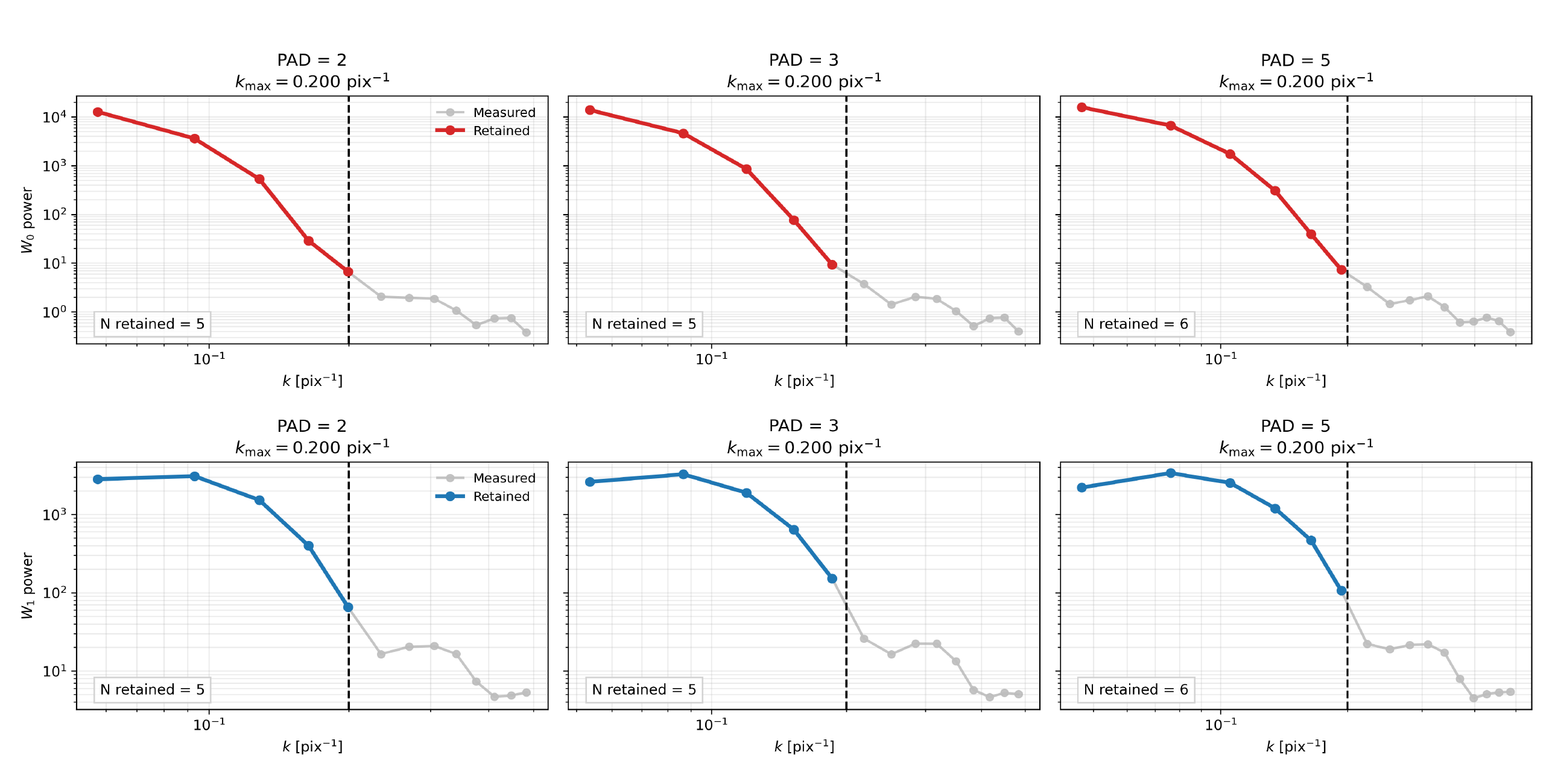}
    \caption{
    Effect of the spatial extraction padding on the power spectra of
    cloud~1177, selected as a representative member of the analysed
    molecular-cloud sample.
    The Fourier cutoff is fixed at one beam
    ($k_{\rm max}=0.2~\mathrm{pix}^{-1}$).
    The top row shows the integrated-intensity ($W_0$) power spectrum,
    while the bottom row shows the first-moment ($W_1$) power spectrum.
    The three columns correspond to $\mathrm{PAD}=2$, 3, and 5 pixels.
    Grey curves show the complete measured power spectra, while the
    coloured segments indicate the Fourier modes retained in the
    calculation of the Brunt-Federrath correction factors.
    The vertical dashed line marks the adopted one-beam cutoff, and the
    number of retained Fourier bins is reported in each panel.
    Increasing the padding extends the sampled spectra towards lower
    spatial frequencies while leaving their shape within the retained
    Fourier range largely unchanged.
    }
    \label{padding_fig}
\end{figure*}

\begin{figure*}
    \centering
    \includegraphics[width=\textwidth]
    {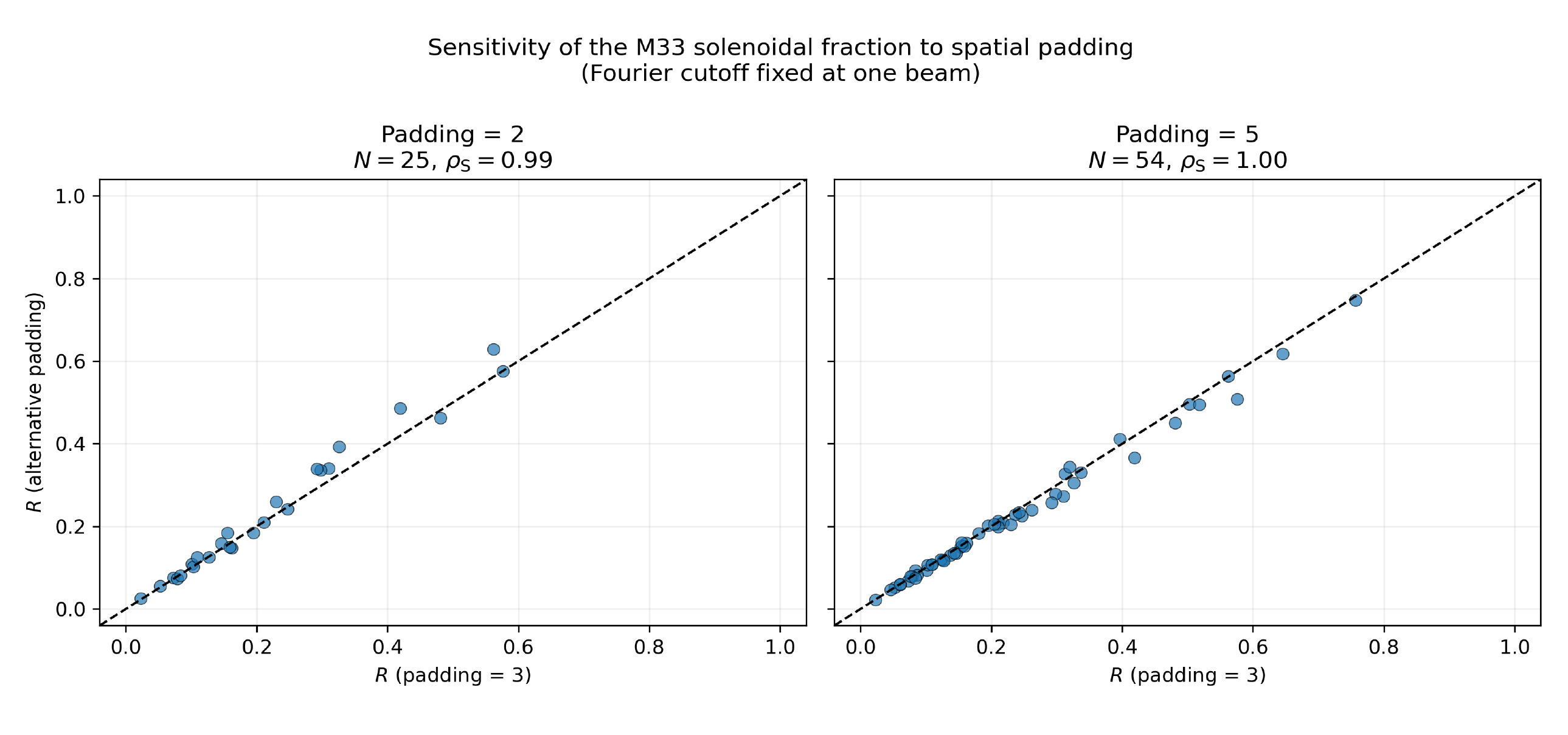}
    \caption{
    Sensitivity of the recovered solenoidal fraction to the spatial
    extraction padding, with the Fourier cutoff fixed at one beam.
    The left panel compares the 25 clouds yielding valid measurements
    with both $\mathrm{PAD}=2$ and $\mathrm{PAD}=3$, while the right
    panel compares the complete 54-cloud sample obtained with
    $\mathrm{PAD}=3$ and $\mathrm{PAD}=5$.
    The dashed lines show the one-to-one relation.
    The close agreement and high Spearman rank coefficients demonstrate
    that the relative ordering and absolute values of the inferred
    solenoidal fractions are insensitive to reasonable changes in the
    extraction padding.
    }
    \label{padding_R_fig}
\end{figure*}

\section{Turbulence analysis}

\subsection{The solenoidal fraction}\label{solenoidal_fraction}

The characterisation of turbulence modes within molecular clouds is
performed using the statistical formalism developed by
\citet{Brunt2010,Brunt2014}, which provides a statistical estimate of
the fraction of turbulent kinetic energy contained in solenoidal
(divergence-free) motions using spectroscopic observations alone.
The method exploits the statistical relationship between the
three-dimensional momentum density field,
$\mathbf{p}=\rho\mathbf{v}$, and its observable line-of-sight projection.
Under the assumptions of statistical isotropy and smoothly vanishing
boundary conditions, the projected momentum density field preserves
statistical information about the solenoidal component of the full
three-dimensional field.
In Fourier space, the line-of-sight projection corresponds to the
$k_z=0$ slice of the three-dimensional momentum density field,
allowing the fraction of power contained in solenoidal modes to be
recovered statistically.
Using Parseval's theorem, the solenoidal fraction, $R$, can be
expressed in terms of the power spectra of observable moment maps
derived from spectral-line data cubes.

In practice, the solenoidal fraction of an individual molecular cloud
is estimated as

\begin{equation}\label{b3}
R = \Bigg[\frac{\langle W_1^2\rangle}{ \langle
W_0^2\rangle}\Bigg]\Bigg[\frac{\langle W_0^2 / \langle W_0
\rangle^2 \rangle}{1 + B_1(\langle W_0^2\rangle/ \langle W_0 \rangle^2 -1)} \Bigg]
\Bigg[ g_{21} \frac{\langle W_2 \rangle}{\langle W_0 \rangle}
\Bigg]^{-1}B_2
\end{equation}

\noindent
with

\begin{equation}\label{b4}
     B_1 = \frac{\sum_{k_x}\sum_{k_y}\sum_{k_z}f(k)-f(0)}{\sum_{k_x}\sum_{k_y}f(k)-f(0)},
\end{equation}

and

\begin{equation}\label{b5}
     B_2 = \frac{\sum_{k_x}\sum_{k_y}\sum_{k_z}f_\perp(k)\frac{k_x^2 + k_y^2}{k^2}}{\sum_{k_x}\sum_{k_y}f _\perp (k)}.  
\end{equation}

\noindent
where $f(k)$ and $f_\perp(k)$ denote the azimuthally averaged power
spectra entering the correction factors $B_1$ and $B_2$,
respectively (notation following \citealt{Orkisz2017}).
The angle brackets denote spatial averages, and
$W_0$, $W_1$, and $W_2$ are the zeroth, first, and second velocity
moments,

\begin{equation}
W_0 = \int I(v)\,dv,\qquad
W_1 = \int v\,I(v)\,dv,\qquad
W_2 = \int v^2 I(v)\,dv,
\end{equation}

where $I(v)$ is the spectral-line intensity as a function of velocity in the frame of reference of the centre of mass of the cloud.

The correction factor $g_{21}$ can be expressed in terms of the density
and velocity fields, the mean density $\rho_0$, and the variance of the
normalized three-dimensional volume density,
$\left\langle (\rho/\rho_0)^2 \right\rangle$.

\begin{equation}\label{b6}
    g_{21} = \frac{\langle\rho^2v^2\rangle/\langle 
    \rho^2\rangle}{\langle \rho v^2\rangle/\langle \rho
    \rangle} =  \Bigg \langle \frac{\rho^2}{\rho_0^2} \Bigg
    \rangle^\epsilon, 
\end{equation}

where $\rho_0$ denotes the mean volume density and $\epsilon$ is a
small positive constant describing the dependence of the velocity
variance on density,

\begin{equation}
\sigma_v^2 \propto \rho^{-\epsilon}.
\end{equation}

In practice,
$\epsilon$ must be estimated empirically from the data, as described
in Section~\ref{density-velocity}. Previous Galactic applications adopted $\epsilon\simeq0.15$ following
the empirical linewidth-density analysis of \citet{Orkisz2017}
and its subsequent application by \citet{Rani2022}. Here we follow
the Brunt-Federrath definition explicitly and determine $\epsilon$
directly from the M33 data by fitting the relation between velocity
variance and cloud density (Section~4.3).

Whether this calibration remains applicable to molecular clouds in an
external galaxy is, however, unknown.
Rather than adopting the Galactic value, we therefore determine
$\epsilon$ directly from the M33 data through the observed
linewidth-density relation (Section~\ref{density-velocity}),
ensuring that the density-velocity correction is calibrated
self-consistently for the extragalactic sample.

\subsection{Assumptions}\label{assumptions}

The Brunt-Federrath formalism described in the previous section relies
on three principal assumptions: (i) the turbulence is statistically
isotropic, (ii) the observed emission decreases smoothly towards the
boundaries of the analysed region, and (iii) the molecular tracer is
optically thin. The first two assumptions are primarily determined by
the spatial resolution of the observations, whereas the optical
thinness of the tracer must be verified explicitly.

The ACA observations of M33 provide a physical resolution of
approximately 30~pc, sufficient to resolve typical giant molecular
clouds in the Milky Way \citep{Solomon1987}, but substantially larger
than the sub-parsec filamentary and clumpy structures commonly observed
within Galactic molecular clouds
\citep[e.g.][]{Arzoumanian2011}. At this resolution, much of the small-scale
anisotropic structure is unresolved and the measured emission is
dominated by cloud-scale morphology \citep{Ossenkopf2002, Cho2003}. Statistical isotropy therefore
provides a reasonable approximation for the ensemble of clouds
analysed here, although individual clouds may still contain resolved
internal structure and need not be morphologically symmetric.

The second assumption concerns the behaviour of the emission at the
boundaries of the Fourier-analysis region. It does not require the
cloud to be internally smooth, but rather that the intensity should
not be truncated abruptly at the edges of the extracted field. Sharp
boundary discontinuities would introduce spectral leakage and
artificial high-spatial-frequency power into the Fourier spectrum.
We address this through the cloud-extraction procedure and investigate
it explicitly in Section~\ref{padding}, where we examine the influence
of different padding sizes on the recovered solenoidal fraction.

The remaining assumption concerns the optical depth of the molecular
tracer. Under the assumption of local thermodynamic equilibrium (LTE),
we derived optical-depth maps for the CO isotopologues using the
publicly available Python package {\sc colte}\footnote{\url{https://github.com/ajrigby/COLTE}},
which solves the radiative transfer equation on a voxel-by-voxel basis
using the observed line intensities. The excitation temperature,
$T_{\rm ex}$, was estimated from the optically thick
$^{12}$CO($2\!-\!1$) emission, assuming that the line is thermalised
and fills the beam. This temperature was then used to derive the
optical depth of the $^{13}$CO($2\!-\!1$) emission through the
standard radiative transfer equation. A further uncertainty arises from the beam-filling assumption.
At the $\sim30$-pc physical resolution of the ACA observations,
individual molecular clouds are only marginally resolved and may
contain substantial unresolved substructure. The assumption that the
optically thick $^{12}$CO emission fully fills the beam may therefore
lead to an underestimate of the true excitation temperature if the
emitting structures occupy only a fraction of the beam. If
$^{12}$CO and $^{13}$CO have similar filling factors, much of this
beam dilution affects the two isotopologues in the same sense.
However, if the $^{13}$CO emission is systematically more compact,
the derived $\tau_{13}$ values may be underestimated. We therefore
regard the optical-depth estimates as approximate and include
beam-filling effects among the systematic uncertainties of the
analysis.

Across the M33 data cube, the derived $^{13}$CO optical depths remain
low, ranging from values consistent with zero to a maximum of $\tau_{13}\simeq0.34$. Optical depths of broadly
similar magnitude have been reported for Galactic molecular clouds
using other $^{13}$CO rotational transitions
\citep{RomanDuval2010,Rigby2016}, although a direct quantitative
comparison is not warranted because the different transitions probe
different excitation conditions. The maximum
value represents only the most optically thick locations in the cube;
typical optical depths within the individual clouds are substantially
lower. Thus, even in the regions of strongest $^{13}$CO emission, the
line remains at most moderately optically thick, while the bulk of the
emission lies firmly in the optically thin regime. Within these systematic uncertainties, the low inferred optical
depths support the use of $^{13}$CO(2--1) as an approximately
optically thin tracer for the present Brunt-Federrath analysis.

These physical assumptions are accompanied by practical observational
requirements that further restrict the cloud population to which the
method can be applied. The emission must be sufficiently well sampled
spatially to retain a useful range of Fourier modes below the adopted
beam cutoff, while projected cloud structures must remain sufficiently
isolated for an unambiguous column-density reconstruction. These
requirements progressively reduce the available sample. The initial
$^{13}$CO extraction contains 175 sources, of which 162 overlap a
$^{12}$CO structure in full PPV space. Removing clouds whose projected
footprints overlap reduces this to a parent sample of 128 clouds, from
which the Fourier-space quality criteria select the primary 54-cloud
sample and the more restrictive 25-cloud subsample discussed in
Section~\ref{sample}. These selections should therefore be regarded as
part of the applicability conditions of the method rather than as a
statistically complete sampling of the molecular-cloud population of
M33.

\subsection{Density-velocity correlations}\label{density-velocity}

To estimate the exponent $\epsilon$ entering the density--velocity
correlation term of the Brunt-Federrath (2014) formalism, we examine
the relation between the observed line-of-sight velocity variance,
$\sigma_{\rm los}^2$, and the mean three-dimensional number density
of the clouds. Following the Brunt--Federrath definition, we fit

\begin{equation}
\sigma_{\rm los}^2 \propto n_{\rm H_2}^{-\epsilon}.
\end{equation}

Under the assumption of isotropy,
$\sigma_{\rm 3D}^2=3\sigma_{\rm los}^2$. Since this conversion is a
constant multiplicative factor for all clouds, it changes only the
normalization of the relation and not its logarithmic slope. We
therefore determine $\epsilon$ directly from the observed
line-of-sight velocity variance.

For each source, the characteristic volume density is computed from
the cloud mass derived from the H$_2$ column-density map and the
intrinsic cloud volume assuming spherical geometry,

\begin{equation}
n_{\rm H_2} =
\frac{3M_{\rm cloud}}
     {4\pi R^3\,\mu_{\rm H_2}m_{\rm H}},
\end{equation}

where $R$ is the beam-deconvolved equivalent radius. We stress that
$n_{\rm H_2}$ therefore represents a cloud-averaged volume density,
rather than a mass- or intensity-weighted density of the
$^{13}$CO-emitting gas. At the $\sim30$-pc spatial resolution of the
observations, unresolved dense substructure occupies only part of the
cloud volume, and the resulting mean densities can consequently be
substantially lower than the characteristic densities locally traced
by $^{13}$CO.
A mass- or intensity-weighted estimate would preferentially emphasise
the denser substructure and would therefore yield larger characteristic
densities. Such a weighting cannot be reconstructed reliably from the
present observations because the internal three-dimensional density
structure is unresolved. The densities used here should thus be
interpreted only as characteristic cloud-scale mean densities for
estimating the empirical linewidth-density relation.

To maintain consistency with \citet{Muraoka2023}, we use the
extrapolated intensity-weighted second moments returned by
{\sc pycprops}. The quantities $\sigma_{\rm maj}$ and
$\sigma_{\rm min}$ represent the rms spatial extents of each cloud
along its major and minor axes, respectively; equivalently, they are
the standard deviations of a Gaussian having the same second moments,
rather than half-widths at half-maximum. After deconvolution of the
spatial moments from the telescope beam ($\sigma_{\rm beam}$), the effective cloud radius is
defined as
\begin{equation}
r_{\rm eff}
 = 1.91
 \left[
 \left(\sigma_{\rm maj}^{2}-\sigma_{\rm beam}^{2}\right)
 \left(\sigma_{\rm min}^{2}-\sigma_{\rm beam}^{2}\right)
 \right]^{1/4}.
\end{equation}

The line-of-sight velocity dispersion, $\sigma_v$, is obtained from
the extrapolated second moment along the spectral axis,
$\sigma_{v,\mathrm{ext}}$, after correcting for the finite spectral
resolution. Following the CPROPS prescription of
\citet{Rosolowsky2006}, a spectral channel of width
$\Delta v_{\rm chan}$ is represented by an equivalent Gaussian
having the same integrated area. Its dispersion is therefore

\begin{equation}
\sigma_{v,\rm chan}
=
\frac{\Delta v_{\rm chan}}{\sqrt{2\pi}}.
\end{equation}

The intrinsic velocity dispersion is then obtained by subtracting
this instrumental contribution in quadrature,

\begin{equation}
\sigma_v =
\left(
\sigma_{v,\mathrm{ext}}^2 -
\frac{\Delta v_{\rm chan}^2}{2\pi}
\right)^{1/2}.
\end{equation}

Requiring reliable beam deconvolution of both spatial and kinematic
properties restricts the sample to clouds that are sufficiently
resolved with respect to the ACA beam. This criterion reduces the
number of clouds with well-defined intrinsic properties to 40,
ensuring that the derived densities and velocity dispersions are not dominated by beam smearing. Figure~\ref{sig_den} shows the resulting relation between the
line-of-sight velocity variance, $\sigma_{\rm los}^{2}$, and
$n_{\rm H_2}$ for the resolved subsample. Following the
\citet{Brunt2014} definition, we fit the relation

\begin{equation}
\sigma_{\rm los}^{2} \propto n_{\rm H_2}^{-\epsilon}.
\end{equation}

A linear fit in logarithmic space yields $\epsilon=0.05$, with a
16th-84th percentile bootstrap interval of
$-0.06<\epsilon<0.16$. The fitted exponent is therefore statistically
consistent with zero. Consistently, a Spearman rank test finds no
significant monotonic correlation between $\sigma_{\rm los}^{2}$ and
$n_{\rm H_2}$ ($\rho=-0.07$, $p=0.69$).

These results indicate no significant dependence of the velocity
variance on the mean cloud density across the nearly three orders of
magnitude spanned by the sample.

At the spatial scales probed here ($\sim 30$ pc), velocity
fluctuations therefore appear largely decoupled from the mean cloud
density.

This behaviour is not inconsistent with classical Larson scaling
relations, which primarily link velocity dispersion to spatial
scale rather than directly to volume density. Our result differs
from the value $\epsilon \simeq 0.15$ inferred for the Orion~B
molecular cloud using multiple tracers sampling different effective
densities \citep{Orkisz2017}. In that study, the density-velocity
relation is derived by comparing tracers probing distinct density
regimes, whereas our analysis measures both density and velocity
dispersion within the same molecular-cloud structures. This
difference in methodology may partly account for the lower value
of $\epsilon$ found in M33.
This suggests that $\epsilon$ may be
method-dependent, or that
different environments and spatial
averaging schemes probe
different aspects of the underlying
density-velocity correlation.
Because $\epsilon$ enters the statistical correction factor
$g_{21}$ in the denominator of Equation~\ref{b3}, changes in the
adopted velocity dispersion-density relation primarily produce a
systematic rescaling of the inferred solenoidal fraction, rather
than qualitatively altering its spatial distribution or radial
trends across M33.

We also note that the interpretation of $\epsilon$ depends on how
the density-velocity relation is probed observationally.
In the formulation of \citet{Brunt2014}, $\epsilon$ represents a
statistical correction describing the coupling between density and
velocity fluctuations within a turbulent medium. Observational
estimates, however, depend on the adopted tracers and spatial
averaging. Multi-tracer approaches, such as that of
\citet{Orkisz2017}, probe density contrasts across different gas
phases, while single-tracer analyses such as the present work
characterise variations within a common molecular component.

For comparison, we also computed $\epsilon$ using projected cloud
radii derived from the footprint area without beam deconvolution.
This alternative approach yields $\epsilon \approx 0.33 \pm 0.23$,
consistent within uncertainties with values measured for Milky Way
molecular clouds \citep{Orkisz2017}, although still statistically
compatible with $\epsilon = 0$ at the $\sim1.5\sigma$ level.

The larger slope is likely to reflect the inclusion of beam-smeared
geometrical effects, which can artificially enhance apparent
density-velocity correlations.
We therefore adopt the value derived from the deconvolved cloud
properties as our fiducial estimate of $\epsilon$, as it is less
affected by beam smearing and more directly reflects intrinsic cloud
properties. The range $\epsilon \in [0,0.33]$ is retained as a
systematic uncertainty and its impact on the derived solenoidal
fractions is assessed in Section~\ref{discussion}.

\subsection{Beam cutoff}\label{cutoff}

The Brunt-Federrath formalism derives the statistical correction factors
$B_1$ and $B_2$ (Equations~\ref{b4} and \ref{b5}) from the azimuthally
averaged power spectra of the observed moment maps. Consequently, an
accurate determination of the spatial power spectrum is essential for a
reliable estimate of the solenoidal fraction. Because the correction
factors are obtained by integrating the observed Fourier power, any
systematic modification of the power spectrum propagates directly into
the inferred value of $R$.
The azimuthal averaging requires a compromise between retaining sufficient
spectral detail and avoiding radial bins that contain few or no sampled
Fourier modes. Following \citet{Orkisz2017}, we therefore adopt
$N_{\rm bin}=S/1.45$ radial bins, where $S$ is the side length, in pixels,
of the square field used for the Fourier transform. This corresponds to a
radial bin width of approximately 1.45 Fourier pixels, slightly larger
than the diagonal separation $\sqrt{2}$ between neighbouring points on
the Cartesian Fourier grid. This choice reduces the occurrence of empty
bins while preserving the shape of the angle-averaged power spectrum.
The same binning prescription is used for all clouds and for all tested
spatial-frequency cutoffs.

Before examining the effects of the Fourier-space cutoff and extraction
padding, we illustrate the moment maps used in the analysis for cloud
1177, selected as a representative molecular cloud from the analysed
sample. Its location within M33 is indicated in
Figure ~\ref{emission}. Figure~\ref{cloud_1177} shows the extracted
integrated-intensity ($W_0$) and velocity-centroid ($W_1$) maps for
the fiducial $\mathrm{PAD}=3$ configuration. The integrated-intensity and velocity-centroid fields provide the
spatial maps from which the Fourier power spectra entering the
Brunt-Federrath analysis are calculated; the velocity-dispersion map
is included to illustrate the internal linewidth structure of the
representative cloud.
 
Cloud 1177 is used consistently in the following sections to
illustrate the effects of the spatial-frequency cutoff, extraction
padding, and instrumental noise on the recovered power spectra.

At spatial frequencies approaching the telescope resolution, however,
the observed power spectrum no longer represents the intrinsic cloud
structure. Beam convolution acts as a low-pass filter, progressively
attenuating small-scale power and artificially steepening the observed
power spectrum. The $^{13}$CO observations have a synthesized beam of
$7.72\times6.86$~arcsec, corresponding to a circularized FWHM of
approximately $7.3$~arcsec. At the adopted image pixel scale, this is
equivalent to approximately five pixels. We therefore approximate the
beam scale as five pixels when defining the Fourier-space cutoff,
corresponding to a maximum spatial frequency of
$k_{\rm max}=1/5=0.20$~pix$^{-1}$. Although \citet{Brunt2014} discuss beam deconvolution as
a possible correction, we find that deconvolution of the ACA data
amplifies high-frequency noise and produces unstable power spectra.
Rather than attempting to recover these poorly constrained Fourier
modes, we adopt a conservative approach in which the power spectra are
evaluated only up to a maximum spatial frequency, excluding modes
dominated by beam attenuation.

To determine an appropriate cutoff, we compare three maximum spatial
frequencies corresponding to one, 1.5 and 2 beam widths. Figure~\ref{cutoff_fig}
illustrates the effect of these choices on the power spectra of a
representative molecular cloud. As the cutoff becomes progressively more
conservative, an increasing fraction of the high-frequency Fourier modes
is excluded from the calculation of the correction factors $B_1$ and
$B_2$. While this reduces the influence of beam convolution, it also
reduces the number of independent Fourier modes available for the
Brunt-Federrath analysis.

The consequences of this trade-off become apparent when the procedure is
applied to the full molecular cloud catalogue. Using a one-beam cutoff,
54 clouds satisfy the quality criteria required for the Brunt-Federrath
analysis. Increasing the cutoff to 1.5 beam widths reduces the sample to
only four clouds, while a two-beam cutoff leaves no cloud with
sufficient Fourier information for a reliable determination of the
solenoidal fraction. The principal effect of adopting increasingly
conservative cutoffs is therefore not a systematic modification of the
derived turbulence mode, but a rapid loss of analysable molecular
clouds.

For the four clouds that remain after adopting the 1.5-beam cutoff, the
derived solenoidal fractions remain highly correlated with those
obtained using the one-beam cutoff ($\rho_{\rm S}=1.0$), indicating that
the relative ordering of the measurements is essentially preserved.
Although the stricter cutoff produces a modest systematic decrease in
the recovered values of $R$, the extremely small number of surviving
clouds prevents any meaningful statistical comparison with the parent
sample. These results indicate that the inferred turbulence mode is not
strongly driven by the highest retained spatial frequencies; rather, the
dominant consequence of adopting a stricter cutoff is the loss of
statistical completeness.

We therefore adopt a cutoff corresponding to one beam throughout the
remainder of this work. This choice provides the best compromise between
restricting the analysis to spatial frequencies that remain minimally
affected by beam convolution and retaining a sufficiently large sample
for statistical investigation. At the $\sim30$\,pc spatial resolution of
the ACA observations, more conservative cutoffs discard nearly all of
the available Fourier information without producing qualitatively
different estimates of the solenoidal fraction.

\subsection{Padding}\label{padding}

The Brunt-Federrath formalism assumes that the observed emission
decreases smoothly towards the boundaries of the analysed region. In
practice, each molecular cloud is extracted from a finite portion of the
observational map. If the extraction boundary lies too close to the
cloud emission, the image may contain abrupt edge discontinuities.
These discontinuities introduce artificial Fourier power through
spectral leakage and may consequently bias the power spectra entering
the correction factors $B_1$ and $B_2$.

\begin{figure*}
	\includegraphics[width=0.8\textwidth]{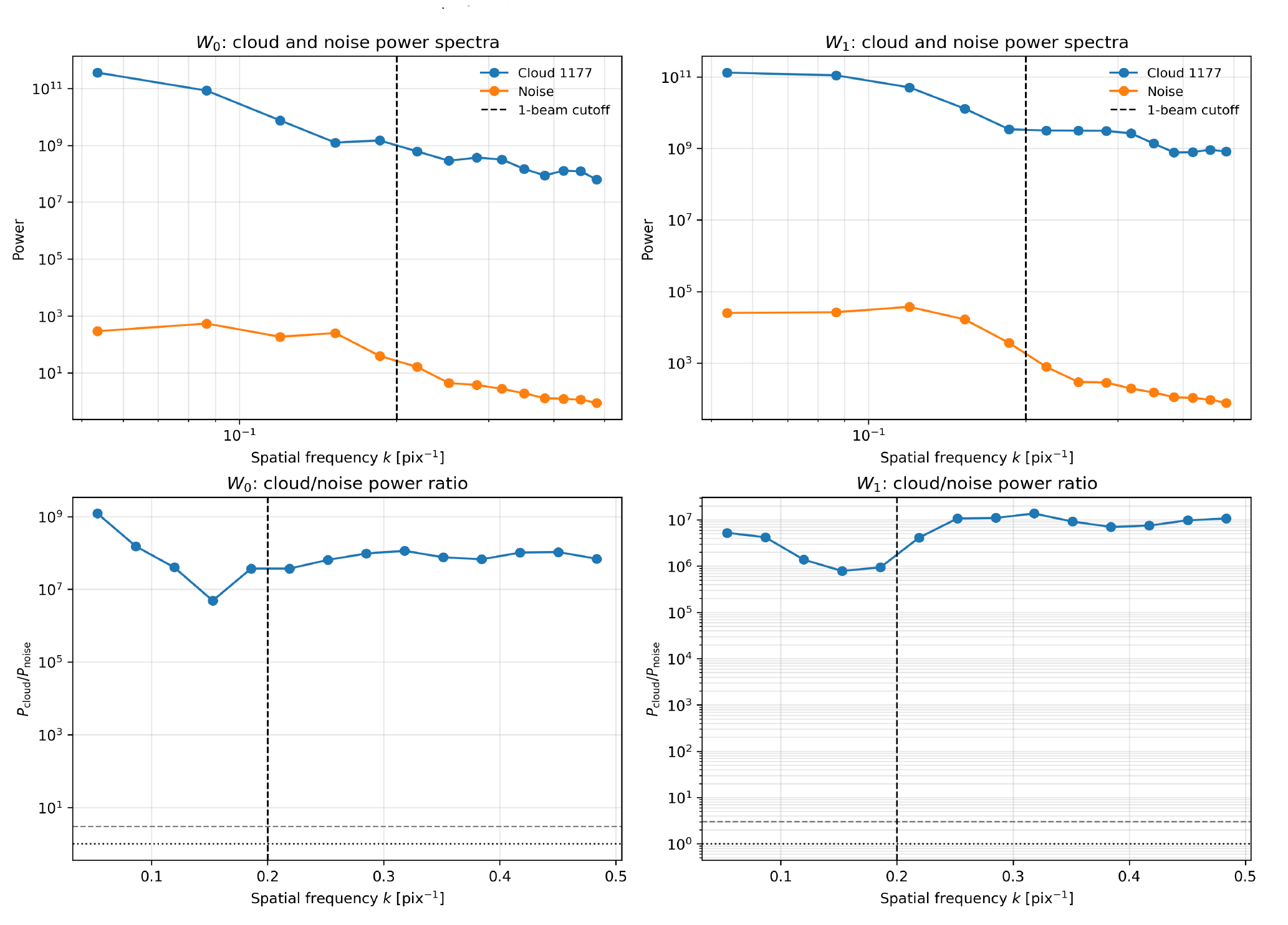}
	\caption{
    Comparison between the power spectra of cloud~1177, selected as a representative molecular cloud from the analysed sample, and a nearby emission-free noise box extracted from the same ACA data cube. The top panels show the one-dimensional power spectra of the integrated intensity ($W_0$; left) and velocity centroid ($W_1$; right), with the vertical dashed line indicating the adopted one-beam spatial-frequency cutoff.
The bottom panels show the corresponding cloud-to-noise power ratios as a function of spatial frequency.
Within the retained Fourier range ($k \leq k_{\rm max}$), the cloud power exceeds the noise power by several orders of magnitude for both $W_0$ and $W_1$, demonstrating that the spatial frequencies used in the Brunt-Federrath analysis are dominated by astrophysical signal rather than instrumental noise.
}
	\label{noise_fig}
\end{figure*}

To reduce this effect, each cloud is embedded within a larger spatial
window by adding a surrounding region around its projected footprint.
The padding changes only the size of the computational domain and does
not alter the emission assigned to the cloud itself. A sufficiently
large extraction window allows the cloud emission to approach the image
boundaries more smoothly, while excessive padding increases the
computational domain without necessarily providing additional
information about the cloud structure.

We investigate the sensitivity of the recovered solenoidal fraction to
three spatial extraction paddings, $\mathrm{PAD}=2$, 3, and 5 pixels,
while keeping the Fourier cutoff fixed at one beam
($k_{\rm max}=0.2~\mathrm{pix}^{-1}$). This isolates the effect of the
finite extraction window from that of the spatial-frequency cutoff
examined in Section~\ref{cutoff}. Figure~\ref{padding_fig} illustrates the effect of varying the extraction padding on the power spectra of cloud 1177, selected as a representative molecular cloud from the analysed sample. Increasing the padding extends the sampled power spectra towards lower spatial frequencies by enlarging the extraction region, while leaving the retained Fourier modes largely unchanged.

The effect of padding on the full cloud sample is shown in
Figure~\ref{padding_R_fig}. For the fiducial choice
$\mathrm{PAD}=3$, 54 clouds yield valid measurements of the solenoidal
fraction. Reducing the padding to two pixels produces a more restrictive
subsample of 25 clouds, whereas increasing the padding to five pixels
retains all 54 clouds. The reduction in sample size at
$\mathrm{PAD}=2$ therefore reflects the smaller extraction regions and
the resulting failure of some clouds to retain sufficient Fourier
information, rather than a systematic change in the cloud population.

For the 25 clouds measured at both $\mathrm{PAD}=2$ and
$\mathrm{PAD}=3$, the recovered solenoidal fractions remain tightly
correlated, with a Spearman rank coefficient
$\rho_{\rm S}=0.99$. The median absolute difference is
$\lvert\Delta R\rvert \simeq 0.011$, corresponding to a median absolute
fractional change of approximately $8$ per cent. The
$\mathrm{PAD}=2$ catalogue is therefore treated as a stricter
quality-controlled subsample rather than as an independent scientific
sample.

Increasing the padding from three to five pixels produces even smaller
changes. All 54 clouds remain analysable, the Spearman rank coefficient
is $\rho_{\rm S}=1.00$, and the median absolute difference is
$\lvert\Delta R\rvert \simeq 0.007$, corresponding to a median absolute
fractional change of approximately $4$ per cent. The mean and median
solenoidal fractions of the two catalogues are also nearly identical.
This demonstrates that the measurements have effectively converged with
respect to further increases in the extraction window.

We therefore adopt $\mathrm{PAD}=3$ throughout the remainder of this
work. Compared with $\mathrm{PAD}=2$, this choice substantially
increases the number of analysable molecular clouds, while the
comparison with $\mathrm{PAD}=5$ demonstrates that further enlargement
of the extraction region produces no meaningful change in the recovered
solenoidal fractions. The adopted padding therefore provides a suitable
balance between satisfying the smooth-boundary requirement of the
Brunt-Federrath formalism and avoiding an unnecessarily enlarged
computational domain.

\subsection{Noise estimation}\label{noise}

The Brunt-Federrath formalism assumes that the observed power spectra
provide a reliable representation of the turbulent structure of the
molecular cloud. If instrumental noise contributes significantly within
the retained Fourier range, the correction factors $B_1$ and $B_2$ may
become biased, leading to systematic errors in the inferred solenoidal
fraction. It is therefore necessary to verify that the adopted
one-beam cutoff excludes Fourier modes where the power spectrum becomes
noise dominated.

To assess the influence of instrumental noise, we compare the power
spectra of a representative molecular cloud (cloud~1177) with those of
an emission-free region extracted from the same ACA data cube. The noise
box is selected to have the same spatial and spectral dimensions as the
cloud cube, ensuring that both regions share identical observational
properties and sampling.

Figure~\ref{noise_fig} compares the one-dimensional power spectra of the
integrated intensity ($W_0$) and first velocity moment ($W_1$) for the
representative molecular cloud and the corresponding noise region. The
bottom panels show the cloud-to-noise power ratios as a function of
spatial frequency. Within the retained Fourier range
($k \leq k_{\rm max}$), the cloud power exceeds the noise power by
several orders of magnitude for both $W_0$ and $W_1$. Only at spatial
frequencies well beyond the adopted one-beam cutoff do the cloud and
noise power spectra begin to approach one another.

The retained Fourier modes are therefore overwhelmingly dominated by
astrophysical signal rather than instrumental noise. The adopted
one-beam cutoff not only excludes spatial frequencies affected by beam
attenuation (Section~\ref{cutoff}), but also remains comfortably within
the regime where the measured power spectra exhibit extremely high
signal-to-noise ratios. Instrumental noise is therefore not expected to
contribute significantly to the uncertainty of the recovered
Brunt-Federrath correction factors or the inferred solenoidal
fractions.

\begin{figure}
	\includegraphics[width=0.5\textwidth]{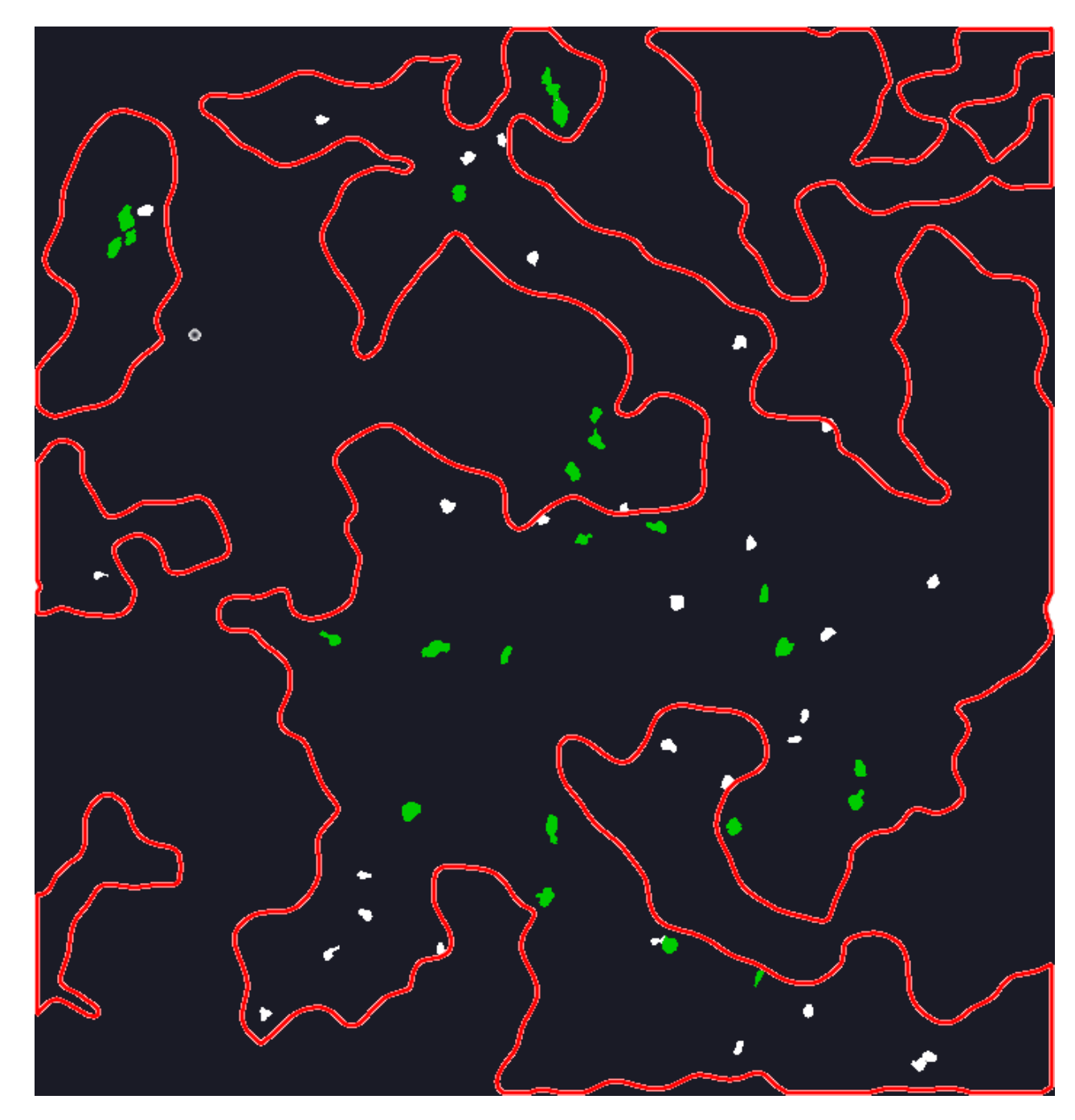}	
\caption{
Spatial distribution of the molecular cloud samples used in this work.
The white regions mark the projected footprints of the 54 molecular clouds selected from the original PYCPROPS catalogue after applying the quality criteria required for the Brunt-Federrath analysis.
The 25 clouds shown in green form the fiducial quality-controlled subsample ($\mathrm{PAD}=2$), used to verify the robustness of the derived solenoidal fractions against a more conservative extraction criterion.
The red contour corresponds to the stellar reference boundary derived from the smoothed \textit{Spitzer}/IRAC $3.6\,\mu$m image (see Figure ~\ref{emission}), tracing the large-scale stellar disc of M33.
}
	\label{segmentation} 
\end{figure}

\subsection{Final sample selection}\label{sample}

The methodological tests presented in the previous subsections establish
that a Fourier cutoff corresponding to one beam provides the best
compromise between minimising the influence of beam convolution and
retaining a statistically meaningful molecular-cloud sample.
Similarly, the padding analysis demonstrates that
$\mathrm{PAD}=3$ yields converged measurements of the solenoidal
fraction while preserving a substantially larger number of clouds than
the more restrictive $\mathrm{PAD}=2$ extraction.

Accordingly, we adopt the catalogue obtained using a one-beam Fourier
cutoff and $\mathrm{PAD}=3$ as the primary scientific sample throughout
the remainder of this work. This catalogue contains 54 molecular clouds
with reliable Brunt-Federrath measurements and forms the basis of all
statistical analyses presented in Sections~\ref{results} and
\ref{discussion}.

For comparison, we also retain the more restrictive
$\mathrm{PAD}=2$ catalogue, comprising 25 molecular clouds.
This subset is not treated as an independent scientific sample, but
rather as a fiducial quality-controlled subsample used to verify that
the principal conclusions remain unchanged under more conservative
selection criteria. Unless otherwise stated, all subsequent analyses are
performed using the 54-cloud primary sample. The Fourier-space quality criteria introduce an expected selection
towards better-sampled clouds. The median number of spatial pixels
increases from $N_{\rm pix}=92.5$ in the 128-cloud parent sample to
174 in the primary 54-cloud sample and 244 in the more restrictive
25-cloud subsample. The corresponding median cloud masses increase
from $1.66\times10^5$ to $2.55\times10^5$ and
$3.42\times10^5\,M_\odot$, respectively. The galactocentric-distance
distributions, however, remain closely comparable, with median values
of 2.55, 2.59, and 2.59~kpc. The final samples are therefore biased
towards larger, better-sampled and more massive structures, but retain
approximately the same radial coverage as the parent catalogue. The spatial distributions of the clouds in the primary and
quality-controlled samples are shown in Figure~\ref{segmentation}.

\begin{figure*}
	\includegraphics[width=1.0\textwidth]{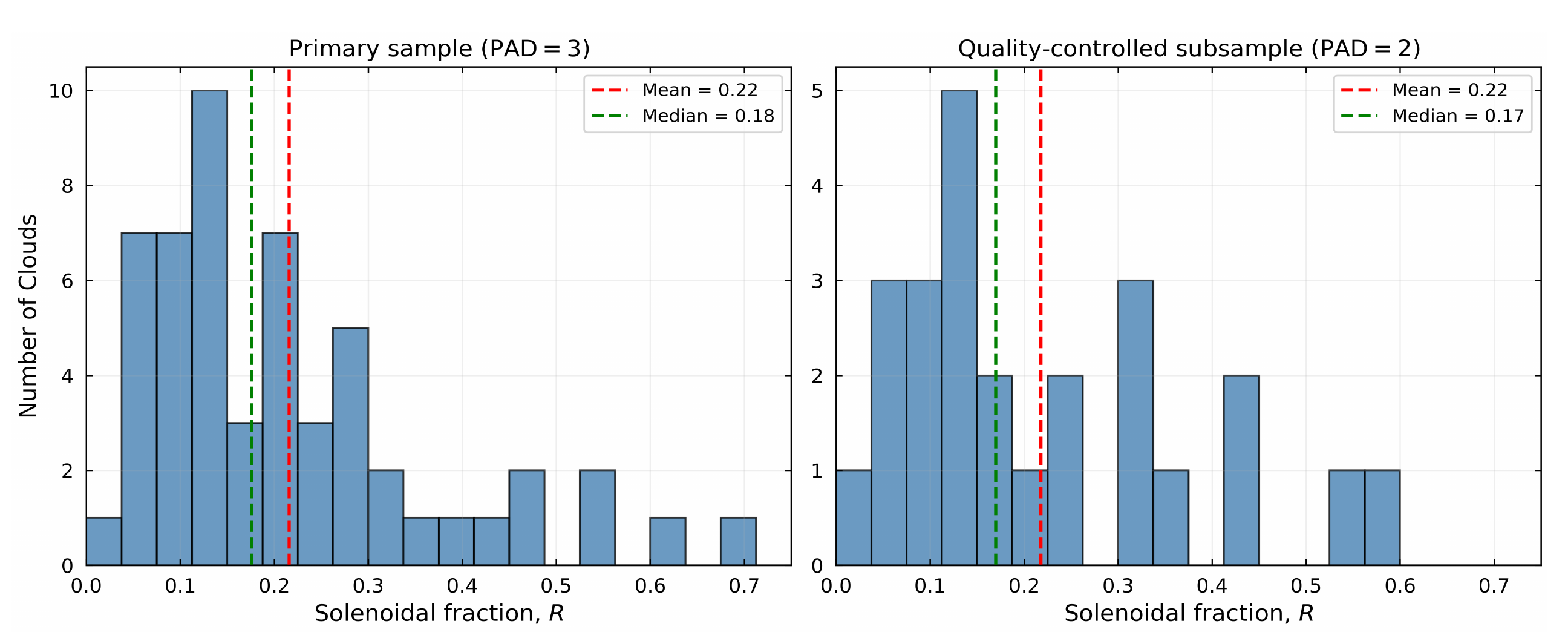}	
	\caption{
Distribution of the measured solenoidal fraction, $R$, for the two molecular cloud samples analysed in this work.
The left panel shows the full sample of 54 molecular clouds extracted using $\mathrm{PAD}=3$, while the right panel shows the higher-quality $\mathrm{PAD}=2$ subsample containing 25 clouds.
The vertical dashed red and green lines indicate the mean and median values of the distributions, respectively.
Both samples exhibit similar distributions, with most clouds characterised by predominantly compressive turbulence, while only a small number of clouds show a dominant solenoidal component.
The close agreement between the two distributions demonstrates that the measured turbulence-mode statistics are robust against the choice of extraction padding.
}
	\label{solenoidal_figure} 
\end{figure*}

The representativeness of the primary sample can also be assessed
relative to the independently segmented ACA+IRAM $^{12}$CO cloud
population. The $^{12}$CO catalogue contains 848 molecular clouds
across the surveyed disc. Because the $^{12}$CO and $^{13}$CO
segmentations were performed independently, the 54 $^{13}$CO clouds
in the primary Brunt-Federrath sample overlap 71 distinct
$^{12}$CO clouds in PPV space. These correspond to only 8.4 per cent
of the $^{12}$CO cloud catalogue by number, but contain approximately
25 per cent of both the total $^{12}$CO cloud luminosity and the
molecular mass represented by the catalogue. The primary sample is
therefore clearly not number-complete, but it contains a substantially
larger fraction of the catalogued molecular-gas budget than its cloud
number alone would suggest. This confirms that the selection
preferentially retains brighter and more massive molecular structures.
Here the quoted fractions refer to emission assigned to catalogued
$^{12}$CO clouds and do not include diffuse $^{12}$CO emission outside
the cloud assignments.

\section{Results}\label{results}

\subsection{Distribution of the solenoidal fraction}\label{distribution}
Having established the robustness of the Brunt-Federrath analysis
and characterised the selection effects associated with the adopted
quality criteria, we now examine the distribution of the recovered
solenoidal fractions within the final M33 molecular-cloud samples.

Figure~\ref{solenoidal_figure} shows the distributions of the measured
solenoidal fraction, $R$, for both the primary molecular-cloud sample
(54 clouds; $\mathrm{PAD}=3$) and the fiducial quality-controlled
subsample (25 clouds; $\mathrm{PAD}=2$). The two distributions exhibit
remarkably similar statistical properties, with mean values of
$R\simeq0.22$ and median values of $R\simeq0.18$ and $R\simeq0.17$,
respectively. The close agreement between the two catalogues confirms
that the inferred turbulence-mode statistics are insensitive to the
adopted extraction padding, consistent with the convergence tests
presented in Section~\ref{padding}.
In both samples, the majority of molecular clouds exhibit
$R<0.5$, indicating that compressive motions dominate the turbulent
momentum-density field. Only a small number of clouds display
predominantly solenoidal turbulence ($R>0.5$), producing a positively
skewed distribution extending towards large values of $R$. No evidence
is found for a bimodal distribution, suggesting instead that the
molecular-cloud population spans a continuous range of turbulence-mode
mixtures dominated by compressive forcing.

\begin{figure*}
	\includegraphics[width=1.0
    \textwidth]{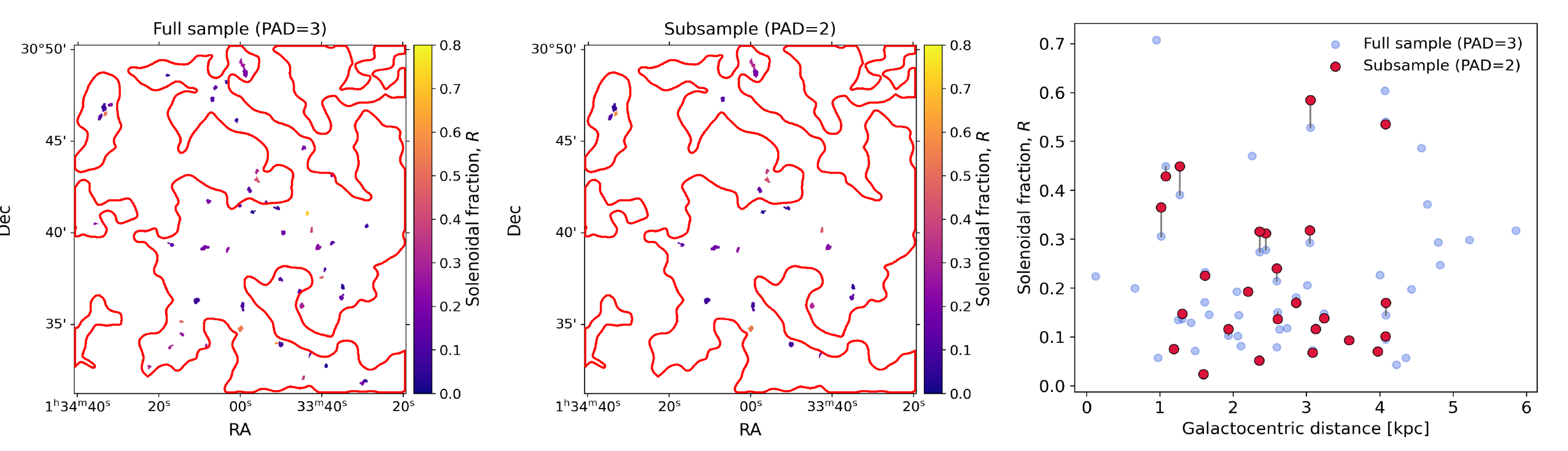}	
\caption{
Spatial distribution of the solenoidal fraction, $R$, across the M33
disc and its dependence on galactocentric distance. The left and middle
panels show the locations of the molecular clouds in the primary
PAD $=3$ sample ($N=54$) and the more restrictive PAD $=2$
quality-controlled subsample ($N=25$), respectively, with each cloud
coloured according to its measured value of $R$. The red contours trace
the adopted spiral-arm structure. The right panel shows $R$ as a
function of deprojected galactocentric distance for the two samples.
The quality-controlled clouds occupy the same radial range and recover
values consistent with those of the primary sample. No clear systematic radial variation is apparent in either sample.
}
	\label{solenoidal_galcen} 
\end{figure*}

The spatial distribution of the solenoidal fraction and its variation
with deprojected galactocentric distance are shown in
Figure ~\ref{solenoidal_galcen}. The primary and quality-controlled
samples display broadly similar spatial behaviour, with low values of
$R$ dominating throughout the observed disc and occasional clouds with
higher solenoidal fractions appearing at several different radii. The
cloud-to-cloud scatter is substantial, and no coherent radial sequence
or preferred galactocentric range is apparent.
The comparable radial coverage indicates that the adopted quality
criteria do not introduce an obvious additional radial bias within the
$^{13}$CO-selected parent sample. This does not, however, remove the
broader sensitivity and mass-selection effects discussed in
Section~\ref{sample}.

These results demonstrate that the global statistical properties of the
solenoidal fraction are robust with respect to the adopted cloud
selection criteria and provide a reliable basis for the environmental
comparisons presented in the following sections.

To assess whether the measured solenoidal fractions are affected by
selection effects related to cloud size or spatial sampling, we examined
the dependence of the basic cloud properties on galactocentric distance,
together with their relation to $R$, for both the primary sample and the
$\mathrm{PAD}=2$ quality-controlled subsample.

For the 54-cloud primary sample, neither the effective cloud radius nor
the number of assigned pixels shows a significant correlation with
galactocentric distance ($\rho=-0.16$, $p=0.26$, and
$\rho=-0.15$, $p=0.28$, respectively).
The corresponding relations are also statistically insignificant for the
25-cloud subsample
($\rho=0.28$, $p=0.18$, and
$\rho=0.20$, $p=0.34$).
The spatial extent and sampling of the analysed clouds therefore do not
vary systematically across the M33 disc.

The solenoidal fraction shows no significant dependence on effective
cloud radius in either the primary sample ($\rho=-0.21$, $p=0.12$)
or the quality-controlled subsample
($\rho=-0.25$, $p=0.22$).
A moderate anti-correlation is, however, observed between $R$ and the
number of assigned pixels, with
$\rho=-0.43$ ($p=1.3\times10^{-3}$)
for the primary sample and
$\rho=-0.43$ ($p=0.030$)
for the subsample.
Clouds with larger projected footprints therefore tend to exhibit lower
values of $R$, although this trend is not accompanied by a systematic
variation in cloud size or pixel sampling with galactocentric distance.

Finally, no significant radial trend in the solenoidal fraction is found
for either sample
($\rho=0.13$, $p=0.35$ for the primary sample and
$\rho=-0.15$, $p=0.47$ for the subsample).
These results indicate that the absence of a systematic radial variation
in $R$ is not caused by radial changes in cloud size, spatial sampling,
or the adopted extraction criteria.

In addition, we assessed the statistical uncertainty associated with the
derived solenoidal fractions using a Monte Carlo error-propagation
scheme. For each molecular cloud, 100 noise realizations were generated
by perturbing the data cube with Gaussian noise corresponding to the
measured rms noise level of the ACA observations
($\sigma_{\rm rms}=0.03$\,K), and the complete Brunt-Federrath analysis
was repeated for each realization.

The resulting Monte Carlo distributions remain centred close to, but
slightly above, the unperturbed measurements, indicating a small
positive estimator bias. Defining the bias as 
$\Delta R=R_{\rm MC,med}-R_0$, its mean value is
$\Delta R=0.0136$ for the primary ($\mathrm{PAD}=3$) sample and
$\Delta R=0.0125$ for the quality-controlled ($\mathrm{PAD}=2$)
subsample. The typical relative uncertainties are modest, with mean
(median) values of 8.0 (7.4) per cent for the primary sample and
6.8 (5.9) per cent for the quality-controlled subsample. Valid
solenoidal-fraction measurements were recovered for all 100
realizations of every cloud in both samples. These uncertainties are
small compared with the observed cloud-to-cloud dispersion in $R$,
confirming that the width of the measured distribution is not dominated
by observational noise.

The cloud with the smallest measured solenoidal fraction
($R\approx0.02$) exhibits a larger fractional uncertainty, as expected,
because even a small absolute uncertainty corresponds to a relatively
large fractional variation. Overall, the Monte Carlo analysis demonstrates
that propagation of the measured 30\,mK channel noise introduces only
modest statistical uncertainties in the recovered solenoidal fractions,
with a small positive estimator bias and typical relative uncertainties
of order ten per cent. The inferred turbulence-mode distribution is
therefore dominated by intrinsic cloud-to-cloud variations rather than
observational noise.

\begin{figure*}
	\includegraphics[width=1\textwidth]{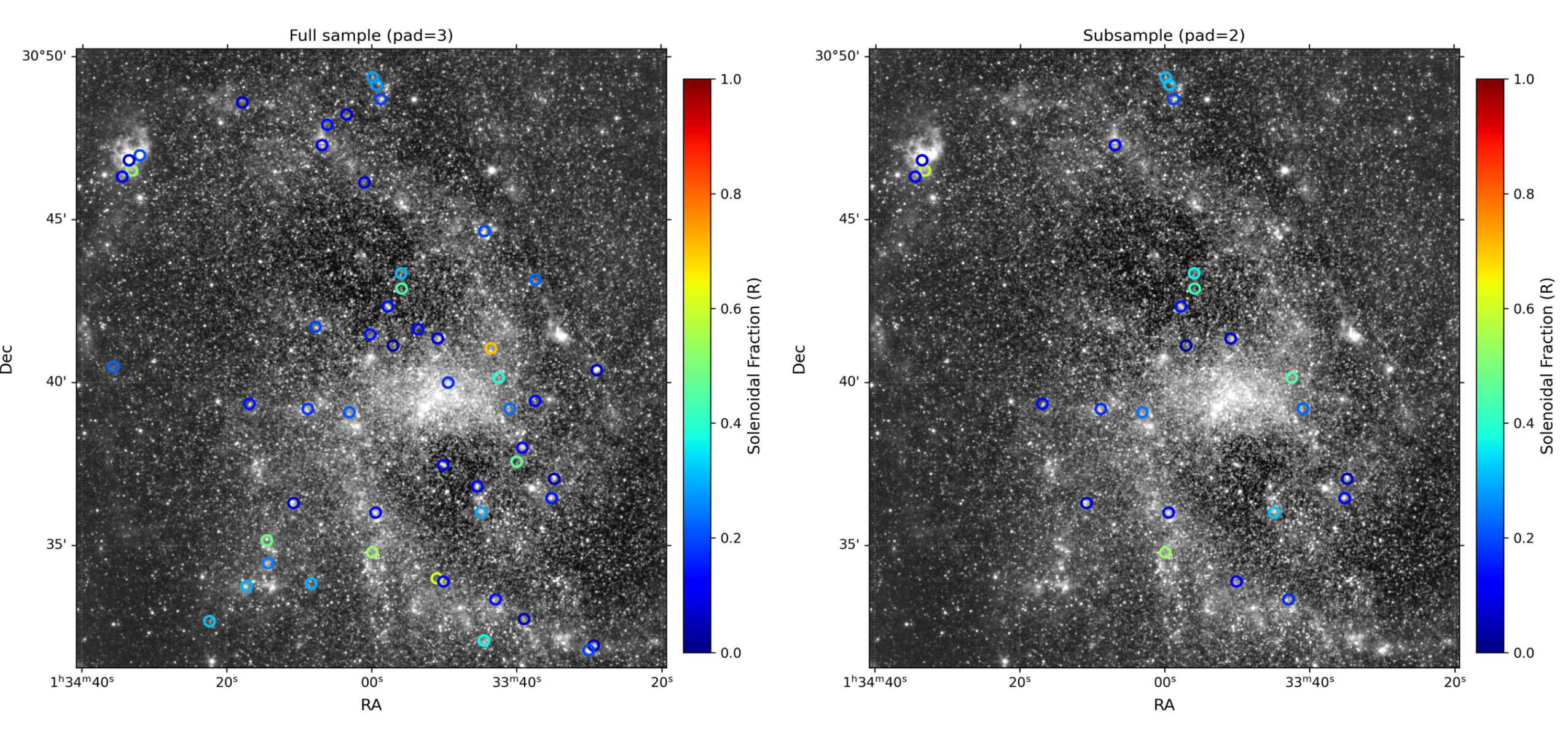}	

\caption{
Spatial distribution of the measured solenoidal fraction, $R$, overlaid on the \textit{Spitzer}/IRAC $3.6\,\mu$m image of M33.
The left panel shows the full sample of 54 molecular clouds ($\mathrm{PAD}=3$), while the right panel shows the higher-quality subsample of 25 clouds ($\mathrm{PAD}=2$).
Each cloud is represented by an open circle coloured according to its measured solenoidal fraction, using the same colour scale in both panels.
The overall distribution of turbulence modes is similar in the two samples, with predominantly compressive clouds ($R<0.5$) found throughout the observed disc and only a small number of clouds exhibiting larger solenoidal fractions. The close correspondence between the two panels demonstrates that the inferred turbulence-mode distribution is robust against the adopted extraction padding.
}
	\label{spiral_arms} 
\end{figure*}

\subsection{Spiral arms}

To investigate whether the measured turbulence modes are related to the
large-scale spiral structure of M33, we compared the spatial
distribution of the molecular-cloud sample with the spiral-arm pattern
traced by the axisymmetric-model-subtracted
\textit{Spitzer}/IRAC 3.6\,$\mu$m image of
\citet{Konishi2024}. By removing the smooth stellar disc, this image
enhances the non-axisymmetric stellar component and provides a
high-contrast representation of the spiral arms.

Figure~\ref{spiral_arms} shows the locations of the molecular clouds,
colour-coded according to their measured solenoidal fractions.
The left panel displays the primary 54-cloud sample, while the right
panel shows the fiducial quality-controlled subsample.
The overall spatial distributions of the two samples are very similar,
again demonstrating that the adopted extraction padding does not alter
the inferred turbulence-mode statistics.

Visual inspection reveals no obvious association between the measured
solenoidal fraction and the spiral-arm morphology.
Clouds with relatively large and small values of $R$ are found
throughout the observed disc, and the few clouds exhibiting relatively
strong solenoidal components are not preferentially concentrated along
the spiral arms, their leading or trailing edges, or other distinct
large-scale stellar structures.
Likewise, predominantly compressive clouds are found both within the
spiral arms and in the inter-arm regions sampled by the survey.
The absence of clear spatial segregation suggests that, at the
$\sim30$\,pc spatial resolution of the present observations, the
balance between compressive and solenoidal turbulent motions is not
primarily regulated by the large-scale spiral structure. This may be
expected for M33, whose flocculent morphology implies a relatively
weak and fragmented spiral-arm potential compared with that of a
grand-design spiral galaxy. The observed cloud-to-cloud variations
instead appear to arise locally, without an obvious correspondence
to the underlying stellar spiral pattern.

We note that the analysed clouds occupy only the regions where significant $^{13}$CO emission is detected in the ACA survey, and the
present sample does not uniformly cover the full optical extent of the
M33 disc. Consequently, subtle environmental variations cannot be ruled
out, but no evidence for a strong spiral-arm dependence is found within
the current molecular-cloud sample.

\begin{figure*}
	\includegraphics[width=1.0\textwidth]{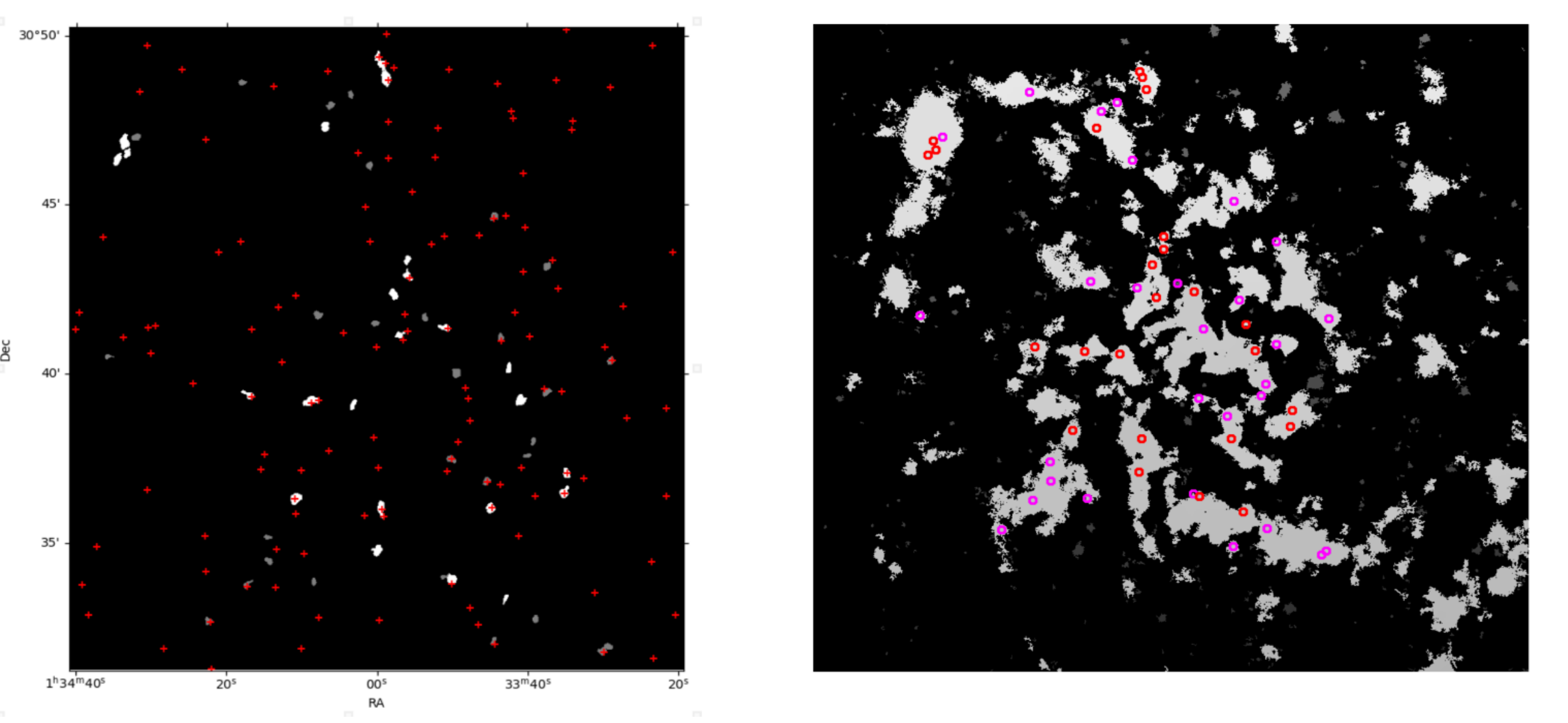}	
\caption{
Association between the molecular cloud samples and the $70\,\mu$m emission in M33.
(a) Positions of the PACS $70\,\mu$m point sources (red crosses) overlaid on the molecular cloud footprints. The grey regions show the 54 molecular clouds in the full ($\mathrm{PAD}=3$) sample, while the white regions indicate the 25-cloud $\mathrm{PAD}=2$ subsample.
(b) Segmentation of the extended PACS $70\,\mu$m emission obtained with \textsc{Photutils}. Magenta circles mark the centroid positions of the full molecular cloud sample, while red circles identify the centroids of the $\mathrm{PAD}=2$ subsample.
}

\label{pacs70} 
\end{figure*}

\begin{figure*}
	\includegraphics[width=1.0\textwidth]{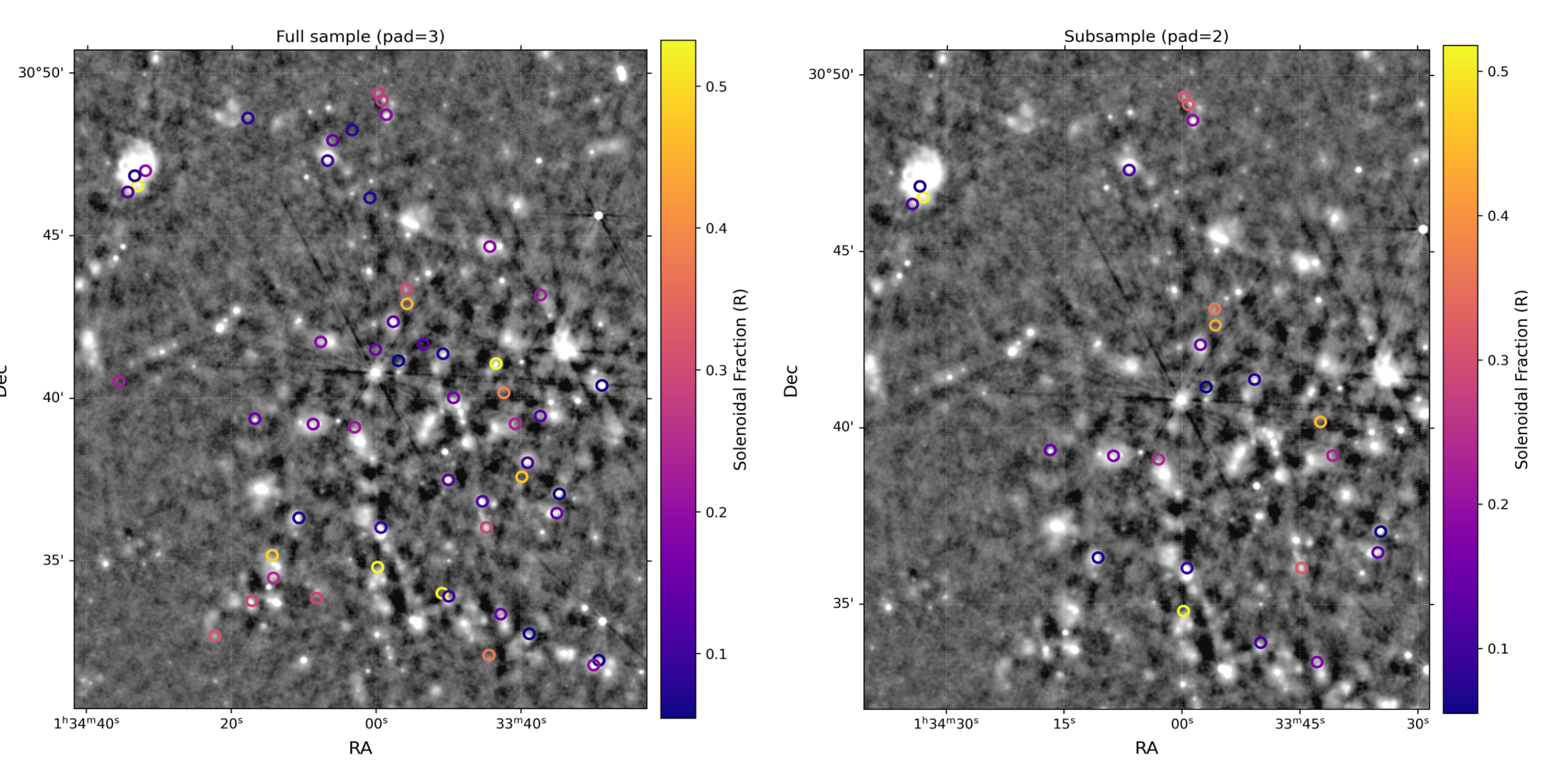}	
	\caption{
Spatial distribution of the measured solenoidal fraction, $R$, overlaid on the VLA 5~GHz radio-continuum map of M33.
The left panel shows the full sample of 54 molecular clouds ($\mathrm{PAD}=3$), while the right panel shows the higher-quality $\mathrm{PAD}=2$ subsample.
Cloud centroids are represented by circles colour-coded according to their measured solenoidal fraction, using the same colour scale in both panels.
The 5-GHz continuum contains both thermal free-free emission from
H{\sc ii} regions and non-thermal synchrotron emission.
}
	\label{5ghz} 
\end{figure*}

\subsection{Relation between solenoidal fraction and star formation tracers}

Previous studies of Galactic molecular clouds have reported a connection
between turbulence properties and star formation activity, with
predominantly compressive turbulence generally associated with enhanced
star formation efficiency
(e.g.~\citealt{Orkisz2017, Rani2022}).
To investigate whether a similar environmental dependence is present in
M33, we compare the molecular-cloud samples with two independent tracers
of recent star formation: the \textit{Herschel}/PACS 70\,$\mu$m emission
from the HerM33ES survey \citep{Kramer2010} and the VLA 5\,GHz radio
continuum.

Figure~\ref{pacs70} compares the molecular-cloud samples with the
70\,$\mu$m emission.
The left panel shows the PACS point-source catalogue overlaid on the
molecular-cloud footprints, while the right panel presents a
\texttt{photutils}-based segmentation of the extended 70\,$\mu$m
emission.
At the spatial resolution of the present observations, the compact PACS
sources rarely coincide directly with the individual
$^{13}$CO cloud footprints identified by {\sc pycprops}.
In contrast, the segmented extended emission follows the distribution of
the molecular clouds much more closely, indicating that the 70\,$\mu$m
emission primarily traces large H\,{\sc ii} region and
photodissociation-region complexes rather than deeply embedded
protostellar sources.
Consequently, unlike Galactic studies where compact 70\,$\mu$m sources
are frequently used to identify very young star-forming objects, the
70\,$\mu$m emission in M33 predominantly traces a somewhat more evolved
phase of massive star formation.

The relation between the measured solenoidal fractions and the radio
continuum is shown in Figure~\ref{5ghz}. The 5-GHz map has an angular
resolution of $5.9$~arcsec FWHM, corresponding to a physical scale of
approximately 24~pc at the distance of M33 \citep{White2019}, and is
therefore closely matched to the spatial resolution of the molecular
cloud data.

At this frequency, the radio continuum contains a substantial thermal
free-free contribution from H\,{\sc ii} regions powered by massive,
short-lived stars, together with non-thermal synchrotron emission from
cosmic-ray electrons accelerated by supernova remnants, stellar winds,
and shocks \citep{Condon1992,Murphy2011,Tabatabaei2007}. The relative
contribution of the thermal component is enhanced compared with that at
lower radio frequencies, making the 5-GHz emission an extinction-free
tracer of recent massive star formation on Myr timescales.

The molecular-cloud population broadly follows the regions of enhanced
radio emission across the observed disc, confirming the close spatial
association between molecular gas and recent massive star formation.
As expected, the brightest radio-continuum complexes also broadly
coincide with regions of enhanced 70-$\mu$m emission, since both are
linked to young massive stars. Their spatial correspondence is not
strictly one-to-one, however, because the infrared emission traces
dust heating, whereas the radio continuum contains both compact
free-free emission and a synchrotron component that can extend beyond
the immediate star-forming regions. Nevertheless, no obvious
segregation is observed between clouds exhibiting different
solenoidal fractions. Clouds spanning the full measured range of $R$
are found both within the brightest radio-continuum complexes and in
regions of comparatively weak radio emission.

Taken together, the PACS and radio-continuum comparisons indicate that,
although the analysed molecular clouds are clearly associated with
active star-forming regions, the measured balance between compressive
and solenoidal turbulent motions does not exhibit an obvious dependence
on these local tracers of recent star formation. Combined with the
absence of significant radial or spiral-arm trends reported in the
previous subsections, these results suggest that, at the
$\sim30$\,pc spatial resolution of the present observations, the
solenoidal fraction is not strongly associated with the present-day
large-scale environment of an individual cloud.

The absence of a spatial correspondence should not, however, be
interpreted as evidence that turbulence mode and star formation are
intrinsically unrelated. At the physical scales probed here, the
infrared and radio emission averages over extended star-forming
complexes and different stages of cloud evolution, potentially
smoothing out the cloud-scale relation between solenoidal fraction and
star formation efficiency identified in Galactic studies
\citep{Rani2022}. A direct quantitative comparison with the Galactic
results will therefore require star formation efficiencies measured
within apertures matched to the individual molecular clouds and at
comparable physical resolution.

\section{Discussion}\label{discussion}

The principal aim of this work was not simply to provide the first measurement of the solenoidal fraction in molecular clouds within M33, but to assess whether the Brunt \& Federrath (2014) statistical framework remains applicable under the observational conditions typical of nearby extragalactic studies. Molecular clouds in M33 are resolved only marginally ($\sim30$\,pc), substantially more coarsely than the Galactic clouds analysed in previous applications of the method \citep{Orkisz2017,Rani2022}. Consequently, the available inertial range is narrower, beam convolution suppresses the highest spatial frequencies, and the assumptions underlying the Fourier-space corrections become more difficult to satisfy. The methodological tests presented in this work demonstrate that these limitations do not prevent a reliable application of the Brunt-Federrath formalism. Systematic investigations of the Fourier cutoff, extraction padding, and Monte Carlo noise propagation all show that the inferred solenoidal fractions remain remarkably stable once appropriate quality criteria are adopted. In particular, the agreement between the primary 54-cloud sample and the more restrictive 25-cloud quality-controlled subsample indicates that the recovered turbulence statistics are insensitive to reasonable variations in the cloud extraction procedure. Likewise, the Monte Carlo analysis shows that observational noise contributes only modest uncertainties, with typical relative errors of order ten per cent, substantially smaller than the intrinsic cloud-to-cloud scatter in the measured solenoidal fractions. These results indicate that the practical limitations imposed by current extragalactic CO observations need not prevent reliable measurements of turbulence modes. Instead, careful treatment of the available spatial frequencies and rigorous quality control are sufficient to obtain reproducible measurements of the solenoidal fraction, even for molecular clouds observed close to the resolution limit. M33 therefore provides a valuable test case for extending the Brunt–Federrath framework beyond the Milky Way.

Having established that the Brunt-Federrath framework can be applied
robustly to the M33 molecular-cloud sample, it is important to assess
whether the measured solenoidal fractions could nevertheless be
systematically affected by the adopted model assumptions. The principal
remaining source of systematic uncertainty is the density-velocity
correlation exponent, $\epsilon$, which enters the $g_{21}$ correction
factor. While Galactic studies typically adopt $\epsilon\simeq0.15$
\citep{Orkisz2017}, we obtain $\epsilon\simeq0.05$, consistent with
zero, for M33.

To assess the effect of this choice, we recomputed the solenoidal
fractions adopting $\epsilon=0.15$. For the primary sample, the mean
decreases from $R=0.22$ to $R=0.17$ and the median from $R=0.18$ to
$R=0.14$. For the quality-controlled subsample, the corresponding
values decrease from a mean of $R=0.22$ to $R=0.17$ and from a median
of $R=0.17$ to $R=0.13$.

The recalculated values remain very tightly correlated with the
original measurements, with Spearman coefficients of $\rho=0.998$ for
both samples. The relations are approximately
$R_{\epsilon=0.15}=0.001+0.750R_{\epsilon=0.05}$ for the primary
sample and
$R_{\epsilon=0.15}=-0.001+0.724R_{\epsilon=0.05}$ for the
quality-controlled subsample. Adopting the Galactic value of
$\epsilon$ therefore produces an approximately multiplicative downward
rescaling of $R$, with a modest compression of the distribution toward
lower values, rather than a qualitative change in its shape or in the
relative ordering of the clouds.

This change is insufficient to reconcile the M33 results with the Milky Way, where typical values are $R \sim 0.37$ (mean), and in fact increases the discrepancy. We therefore conclude that uncertainties in $\epsilon$ do not account for the lower values of $R$ observed in M33. Adopting the Galactic value of $\epsilon$  changes the absolute scale of the inferred solenoidal fractions but does not alter the principal conclusions of this work. The M33 cloud population remains systematically shifted toward lower values of $R$ than Galactic molecular clouds, indicating that the observed offset cannot be explained solely by the adopted density-velocity correlation correction.

With the robustness of the methodology established, we can now consider the physical implications of the measured turbulence statistics. The molecular-cloud population in M33 is characterised by a mean solenoidal fraction of approximately $R\simeq0.22$, indicating
that compressive motions dominate the momentum-density field throughout the analysed sample. These values remain systematically lower than those reported for Galactic molecular clouds analysed using the same formalism \citep{Rani2022}, suggesting that molecular clouds in M33 are characterized by a systematically larger contribution from compressive motions than their Galactic counterparts.

The contrast between the values of $R$ found in our sample and Galactic molecular clouds is likely to be linked to the global dynamical state of the host galaxy. M33 is a lower-mass, smaller late-type spiral with a shallower gravitational potential and no prominent stellar bar. In contrast to the Milky Way, where the bar and overall mass distribution drive strong large-scale shear and organized gas flows, the interstellar medium in M33 is distributed in a more weakly bound, flocculent disc with widespread but less structured star formation \citep{Gratier2010,Druard2014}. Molecular clouds in M33 are therefore likely to form within a more dynamically evolving, atomic-dominated environment, where large-scale converging flows and gravitational instabilities play a significant role. These conditions are expected to favour compressive motions because cloud formation driven by large-scale gas accumulation and gravitational collapse naturally enhances density compression within the turbulent flow \citep{Federrath2010}, providing a natural explanation for the systematically lower values of $R$. 

A further indication that the lower solenoidal fractions measured in
M33 may be physically meaningful is provided by its star formation
properties. Several studies have inferred a relatively high
molecular-gas star formation efficiency in M33 compared with more
massive spiral galaxies
\citep{Gardan2007,Gratier2010,Kang2012}. The precise value remains
uncertain, however, because the inferred molecular-gas depletion time
depends on the adopted star formation tracer, CO-to-H$_2$ conversion
factor, and spatial scale. In particular, \citet{Schruba2010} obtained
a median depletion time of approximately 1~Gyr on kiloparsec scales
and showed that the inferred value becomes strongly dependent on
aperture size and target selection below approximately 300~pc.

A comparatively high molecular-gas star formation efficiency would
be qualitatively consistent with the negative relation between
solenoidal fraction and star formation efficiency found for Galactic
molecular clouds by \citet{Rani2022}, in which clouds with a larger
compressive contribution exhibit systematically higher star formation
efficiencies. The lower characteristic values of $R$ measured in M33
may therefore be compatible with the broader picture of a galaxy in
which molecular gas is converted into stars relatively efficiently.
This comparison should nevertheless be regarded as suggestive rather
than quantitative, because the published depletion times are generally
measured on galactic or kiloparsec scales, whereas the present values
of $R$ are derived for individual molecular clouds.

At the scale of individual clouds within M33, however, the present
data do not reveal a direct relation between turbulence mode and
recent star formation. Our comparison with 70-$\mu$m emission traces
only the relative locations of molecular clouds and extended
H\,{\sc ii}/PDR complexes, and therefore does not provide a direct
cloud-by-cloud measurement of the relation between $R$ and star
formation efficiency. No clear spatial correspondence is observed
between the solenoidal fraction and the far-infrared emission, which
traces recent star formation over Myr timescales
\citep{KennicuttEvans2012}. The absence of a clear spatial correspondence should not, however, be
interpreted as evidence that turbulence mode and star formation are
intrinsically unrelated. At the physical scales probed here, the
infrared and radio-continuum measurements blend emission from extended
star-forming complexes and from regions at different evolutionary
stages, potentially washing out the cloud-scale relation between
solenoidal fraction and star formation efficiency identified in the
Milky Way \citep{Rani2022}. A direct comparison with the Galactic
results is therefore difficult at the present resolution and will
require star formation efficiencies measured within apertures matched
to individual molecular clouds. Future infrared observations at
substantially higher angular resolution may enable such a comparison
on physical scales more closely matched to those reached by ALMA.

Similarly, clouds with higher or lower values of $R$ do not occupy
preferred locations relative to the spiral pattern, indicating that
the spiral arms do not imprint a unique turbulence mode on individual
clouds at the spatial scales probed here. The absence of strong
spatial or radial trends may indicate that the turbulent state of a
cloud is determined primarily by local evolutionary processes and
rapidly varying driving mechanisms, rather than retaining a simple
imprint of the large-scale environment in which the cloud formed
\citep{MacLowKlessen2004,Padoan2014}. This behaviour is broadly
consistent with that observed in the Milky Way, where no clear
spiral-arm dependence of the solenoidal fraction was identified
\citep{Rani2022}.

However, a key difference emerges: in the Galactic case, enhanced solenoidal fractions are observed near the ends of the central bar, whereas M33, which lacks a prominent bar, shows no comparable feature. This contrast suggests that large-scale dynamical structures such as bars may influence the baseline balance between compressive and solenoidal motions.

Taken together, the absence of statistically significant radial,
spiral-arm, or star-formation trends suggests that, at the
$\sim30$~pc resolution of the present observations, the balance
between compressive and solenoidal turbulent motions is not strongly
regulated by the present-day large-scale galactic environment.
Instead, the cloud-to-cloud variations in $R$ may primarily reflect
local evolutionary processes or stochastic differences in the
turbulent driving mechanisms.

One important limitation of the present sample is that the $^{13}$CO
emission is not detected uniformly across the full extent of the M33
disc because of sensitivity limitations \citep{Muraoka2023}. The
present sample is therefore biased toward relatively massive molecular
clouds located in regions of enhanced molecular gas and ongoing star
formation, particularly along the spiral structure traced by the
70-$\mu$m emission. This observational bias should be taken into
account when interpreting the absence of strong environmental trends
in the solenoidal fraction.
The extent of this selection can be quantified by comparison with the
independently segmented ACA+IRAM $^{12}$CO cloud population. The
$^{12}$CO catalogue contains 848 molecular clouds across the surveyed
disc. Because the $^{12}$CO and $^{13}$CO segmentations were performed
independently, the 54 $^{13}$CO clouds in the primary Brunt-Federrath
sample overlap 71 distinct $^{12}$CO clouds in PPV space. Because the $^{12}$CO and $^{13}$CO segmentations were performed
independently, their cloud assignments are not constrained to have a
one-to-one correspondence, and individual $^{13}$CO clouds may overlap
more than one $^{12}$CO structure in PPV space. The 54 $^{13}$CO
clouds in the primary Brunt-Federrath sample therefore overlap 71
distinct $^{12}$CO clouds. These
correspond to only 8.4 per cent of the $^{12}$CO cloud catalogue by
number, but contain approximately 25 per cent of both the total
$^{12}$CO cloud luminosity and the molecular mass represented by the
catalogue. The primary sample is therefore not number-complete, but it
contains a substantially larger fraction of the catalogued
molecular-gas budget than its cloud number alone would suggest. The
quoted fractions refer to emission assigned to catalogued $^{12}$CO
clouds and do not include diffuse $^{12}$CO emission outside the cloud
assignments.
A further limitation concerns the physical interpretation of the recovered solenoidal fraction. The Brunt-Federrath formalism
decomposes the observed momentum-density field into solenoidal and
compressive components, but it does not by itself distinguish
stochastic turbulence from coherent cloud-scale motions. Ordered
rotation or shear can therefore contribute to the inferred solenoidal
component, whereas gravitational infall, global contraction,
expansion, and other converging or diverging flows can contribute to
the compressive component. Such motions are physically distinct from
stochastic turbulence but enter the observed velocity field and hence
the Brunt-Federrath decomposition. At the $\sim30$-pc resolution of
the present observations, separating these contributions robustly on
a cloud-by-cloud basis would require additional kinematic modelling
and cannot be achieved from the present data alone. The measured
values of $R$ should therefore be interpreted as characterising the
relative solenoidal and compressive content of the observed
cloud-scale momentum-density field under the turbulence
interpretation, rather than as a decomposition of strictly stochastic
turbulent motions. These limitations are particularly relevant when searching for weak
environmental trends. Although the primary and quality-controlled
samples retain approximately the same galactocentric-distance coverage
as the 128-cloud parent sample, this does not remove the underlying
selection towards $^{13}$CO-bright, larger, better-sampled and more
massive clouds. Furthermore, any environmental dependence in the
relative importance of coherent rotation, collapse, feedback-driven
expansion, or other ordered motions could contribute to the measured
cloud-to-cloud variation in $R$. The absence of statistically
significant radial, spiral-arm, or star-formation trends found here
should therefore be interpreted as applying to the cloud population
and spatial scales sampled by the present observations; weaker trends
in the complete molecular-cloud population of M33 cannot be excluded.

Despite these limitations, the present work demonstrates that the Brunt–Federrath formalism can be applied successfully to nearby galaxies observed close to the practical resolution limit of current interferometric CO surveys. As ALMA observations continue to improve in both sensitivity and spatial resolution, the methodology developed here can be extended to larger samples spanning a wider range of galactic environments, enabling systematic comparative studies of turbulence modes across the nearby Universe.

\section{Conclusions}

We have presented the first application of the Brunt \& Federrath
(2014) statistical framework to molecular clouds in an external galaxy,
using a sample of $^{13}$CO(2-1) clouds identified in the ALMA ACA
survey of M33. Rather than focusing solely on the resulting turbulence
measurements, this work has assessed the applicability of the
Brunt-Federrath formalism under the challenging observational
conditions typical of nearby extragalactic studies, where molecular
clouds are only marginally resolved.

Our principal conclusions are as follows.

\begin{enumerate}

\item The Brunt-Federrath method remains robust when applied to
molecular clouds observed at a physical resolution of
$\sim30$\,pc, provided that appropriate quality criteria are adopted.
Systematic tests of the Fourier-space cutoff, extraction padding and
noise propagation demonstrate that the recovered solenoidal fractions
are stable against these methodological choices.

\item A one-beam Fourier cutoff provides an optimal compromise between
retaining a statistically meaningful cloud sample and minimizing the
effects of beam attenuation on the measured power spectra. Likewise,
the adopted extraction padding ($\mathrm{PAD}=3$) produces converged
results while preserving a substantially larger sample than more
restrictive extraction criteria. The agreement between the primary
54-cloud sample and the 25-cloud quality-controlled subsample confirms
the robustness of the derived turbulence statistics.

\item The molecular-cloud sample analysed here is dominated by
predominantly compressive motions with a mean solenoidal fraction
of approximately $R\simeq0.22$. These values remain systematically
lower than those reported for Galactic molecular clouds analysed using
the same methodology, suggesting that molecular clouds in M33 occupy a
different region of turbulence-parameter space.

\item No statistically significant dependence of the solenoidal
fraction is found on galactocentric distance, spiral-arm morphology,
or the adopted tracers of recent star formation. At the spatial scales
probed by the present observations, cloud-to-cloud variations in the
balance between compressive and solenoidal turbulent motions therefore
appear to dominate over large-scale environmental effects.

\item Monte Carlo propagation of the measured observational noise
introduces only modest uncertainties in the inferred solenoidal
fractions, with typical relative uncertainties of order ten per cent
and only a small positive estimator bias. The observed scatter in the
solenoidal fraction is therefore dominated by intrinsic variations
between molecular clouds rather than measurement uncertainties.

\end{enumerate}

To our knowledge, this work represents the first application of the
complete Brunt-Federrath (2014) formalism to recover the relative
fractions of solenoidal and compressive turbulent motions in
molecular clouds within an external galaxy. While related
extragalactic studies have applied statistical techniques based on the
density-reconstruction method of \citet{Brunt2010} or have inferred
turbulence properties through density probability distribution
functions and multi-line analyses \citep[e.g.][]{Sharda2018,Menon2021,Gerrard2023},
the BF14 momentum-decomposition formalism itself has not
previously been applied to extragalactic molecular-cloud
observations. Our results therefore demonstrate that the method
remains robust at the spatial resolutions currently achievable with
interferometric CO observations of nearby galaxies, opening the way
for future comparative studies of turbulence modes across diverse
galactic environments.
The methodology presented here provides a practical framework for
applying the Brunt-Federrath formalism to future ALMA observations of
nearby galaxies. As increasingly sensitive and higher-resolution
molecular-cloud surveys become available, the approach developed in
this work will enable systematic comparisons of turbulence modes across a broad range of galactic environments, providing new observational
constraints on the role of turbulence in molecular-cloud evolution and
star formation.


\section*{Data Availability}

The ACA $^{13}$CO(2-1), C$^{18}$O(2-1), and $^{12}$CO(2-1) ACA/IRAM maps are taken from \citet{Muraoka2023}. 
The original IRAC 3.6~$\mu$m map is presented by
\citet{Dale2009}, while the axisymmetric-model subtracted map
used to trace the spiral structure is taken from
\citet{Konishi2024}.
The 5~GHz radio continuum data for M33 were taken from the deep VLA survey presented by \citet{White2019}, with data products publicly available at \url{http://sundog.stsci.edu/m33/}. 
The $^{13}$CO(2-1) catalogue constructed for the analysis presented in this article is available for download from the CANFAR archive\footnote{\url{https://www.canfar.net/storage/list/AstroDataCitationDOI/CISTI.CANFAR/26.0010/data}}.



\bibliographystyle{mnras}
\bibliography{example} 





\bsp	
\label{lastpage}
\end{document}